\documentclass[11pt,a4paper,notitlepage]{article}
\usepackage[utf8]{inputenc}
\usepackage[english]{babel}
\usepackage[T1]{fontenc}
\usepackage{hyperref}
\usepackage{amsmath}
\usepackage{amsfonts}
\usepackage{color} 
\usepackage{caption}
\usepackage{array,multirow,makecell}
\setcellgapes{1pt}
\makegapedcells
\newcolumntype{R}[1]{>{\raggedleft\arraybackslash }b{#1}}
\newcolumntype{L}[1]{>{\raggedright\arraybackslash }b{#1}}
\newcolumntype{C}[1]{>{\centering\arraybackslash }b{#1}}
\newcommand{\Tr}{\mathrm{Tr}}
\newcommand{\tr}{\mathrm{tr}}

\newcommand{\sym}{\mathrm{Sym}}

\usepackage{enumerate}
\usepackage{subfig}
\usepackage{young}

\newcommand{\cA}{{\mathcal A}}

\newcommand{\cC}{{\mathcal C}}

\newcommand{\cF}{{\mathcal F}}

\newcommand{\cK}{{\mathcal K}}

\newcommand{\cM}{{\mathcal M}}
\newcommand{\cN}{{\mathcal N}}
\newcommand{\cO}{{\mathcal O}}

\newcommand{\cQ}{{\mathcal Q}}

\newcommand{\cS}{{\mathcal S}}

\newcommand{\cV}{{\mathcal V}}

\newcommand{\bP}{\mathrm{P}}

\definecolor{mygray}{gray}{0.3}

\newcommand\beq{\begin{equation}}
\newcommand\eeq{\end{equation}}
\newcommand{\bes}{\begin{eqnarray}}
\newcommand{\ees}{\end{eqnarray}}
\def\nn{{\nonumber}}
\newcommand{\one}{\mbox{$1 \hspace{-1.0mm}  {\bf l}$}}

\def\psib{\bar{{\psi}}}

\newcommand\restr[2]{{
  \left.\kern-\nulldelimiterspace 
  #1 
  \vphantom{\big|} 
  \right|_{#2} 
  }}
\def\extd{\mathrm {d}}
\newcommand{\U}{\mathrm{U}}
\newcommand{\mO}{\mathrm{O}}
\newcommand{\SU}{\mathrm{SU}}
\newcommand{\Sp}{\mathrm{Sp}}

\newcommand{\usp}{\mathfrak{sp}}

\usepackage{multicol}
\usepackage{amssymb}
\usepackage{graphicx}
\usepackage[left=2cm,right=2cm,top=2.5cm,bottom=2.5cm]{geometry}

\def\le{\lambda_{\mathrm{eff}}^2}

\def\sym{\vcenter{\hbox{\begin{Young}  1  &  2 & 3 \cr \end{Young}}}}
\def\asym{\vcenter{\hbox{\begin{Young} 1  \cr  2 \cr 3 \cr \end{Young}}}}
\def\mixedb{\vcenter{\hbox{\begin{Young}  1  &  3 \cr 2 \cr \end{Young}}}}
\def\mixed{\vcenter{\hbox{\begin{Young}  1  &  2 \cr 3 \cr \end{Young}}}}

\def\mixednn{\vcenter{\hbox{\begin{Young}    &   \cr  \cr \end{Young}}}}

\begin{document}
\begin{center}
\textbf{\Large{{\centering SYK-like tensor quantum mechanics 
with $\Sp(N)$ symmetry}}}
\vspace{15pt}

{\large Sylvain Carrozza$^{a,}$\footnote{\url{scarrozza@perimeterinstitute.ca}}, \large Victor Pozsgay$^{b,}$\footnote{\url{victor.pozsgay@ens-lyon.fr }}}

\vspace{10pt}

{$^a$\sl Perimeter Institute for Theoretical Physics\\
 31 Caroline St N, Waterloo, ON N2L 2Y5, Canada\\
}

\medskip

{$^b$\sl \'{E}cole Normale Sup\'{e}rieure de Lyon\\
 65 All\'{e}e d'Italie, 69007 Lyon, France\\
}

\end{center}

\vspace{5pt}

\begin{abstract}
\noindent We introduce a family of tensor quantum-mechanical models based on irreducible rank-$3$ representations of $\Sp(N)$. In contrast to irreducible tensor models with $\mO(N)$ symmetry, the fermionic tetrahedral interaction does not vanish and can therefore support a melonic large $N$ limit. The strongly-coupled regime has a very analogous structure as in the complex SYK model or in $\U(N) \times \mO(N) \times \U(N)$ tensor quantum mechanics, the main difference being that the states are now singlets under $\Sp(N)$. We introduce character formulas that enumerate such singlets as a function of $N$, and compute their first values. We conclude with an explicit numerical diagonalization of the Hamiltonian in two simple examples: the symmetric model at $N=1$, and the antisymmetric traceless model at $N=3$.
\end{abstract}

\setcounter{tocdepth}{2}
\tableofcontents

\section{Introduction}

The large $N$ limit of tensor models has found applications in a growing list of different subjects over the years. It was initially discovered in the context of discrete approaches to quantum gravity in dimension $d\geq 3$ \cite{expansion1, expansion2, expansion3, critical, RTM}, where it triggered a number of developments, on topics such as: combinatorial aspects and refinements of the large $N$ expansion \cite{uncoloring, expansion4, expansioin5, expansioin6, Carrozza:2015adg, thesisLuca, Bonzom:2018btd}, probability and random geometry \cite{universality, melbp}, non-local field theories \cite{BenGeloun:2011rc, BenGeloun:2012pu, Samary:2014oya, Rivasseau:2017xbk}, group field theory \cite{Carrozza:2012uv, tt2, Samary:2012bw, thesis, Carrozza:2014rba}, statistical physics \cite{IsingD, bonzom2013universality, Delepouve:2015nia}, or functional renormalization group methods \cite{Benedetti:2014qsa, Benedetti:2015yaa, Eichhorn:2017xhy}. In this line of thought, research efforts have been primarily focused on a particular brand of tensor models -- going by the name of \emph{colored} \cite{color} and \emph{uncolored} \cite{uncoloring} tensor models
 -- because of their nice relationship to combinatorial topology and simplicial geometry \cite{lost, review, Casali:2017tfh}. 

More recently, tensor models have found very interesting applications in the more familiar context of quantum mechanics and local quantum field theory. Motivated by the SYK model \cite{Sachdev:1992fk, Kitaev, Maldacena:2016hyu, Polchinski:2016xgd, Gross:2017aos, Rosenhaus:2018dtp}, Witten \cite{Witten:2016iux} and Klebanov-Tarnopolsky \cite{Klebanov:2016xxf} have introduced tensor quantum-mechanical models which develop an emergent conformal symmetry in a suitable large $N$ and strong-coupling regime. These two models have since then been investigated in detail and generalized in a number of directions. These include works on: properties of the large $N$ expansion \cite{Gurau:2016lzk, Gurau:2017qna, Bonzom:2017pqs}, properties of the spectrum (including at small $N$) \cite{Bulycheva:2017ilt, Klebanov:2018nfp, Pakrouski:2018jcc, Krishnan:2016bvg, Krishnan:2017ztz, Krishnan:2017txw, Krishnan:2018hhu}, the infrared structure of such theories \cite{Choudhury:2017tax, Benedetti:2018goh}, generalizations to $d\geq 2$ \cite{Giombi:2017dtl, Prakash:2017hwq, Benedetti:2017fmp, Giombi:2018qgp}, multi-matrix models with similar properties \cite{Ferrari:2017ryl, Azeyanagi:2017drg, Ferrari:2017jgw, Azeyanagi:2017mre}, and connections to higher-spin theories \cite{Vasiliev:2018zer}. We refer to the recent TASI lectures \cite{TASI_tensor-review} for a more exhaustive list of current research topics. The key feature of the large $N$ limit underlying these recent developments is that it is generically dominated by melon diagrams \cite{critical, Carrozza:2015adg, Bonzom:2018jfo}: this family of Feynman graphs turns out to be tractable enough to be of practical use, and rich enough to capture the characteristic bilocal effects of SYK-like strongly coupled phases. On the other hand, colors (better referred to as \emph{flavors} in these examples) do not seem to play any fundamental role in this context. This has motivated new work extending the domain of validity of the large $N$ expansion, from models in which no symmetry at all is assumed among the indices of the tensors -- as is for instance the case in rank-$3$ tensor models with $\mO(N)^3$ symmetry \cite{Carrozza:2015adg, Klebanov:2016xxf} --, to models based on irreducible rank-$3$ tensor representations \cite{Klebanov:2017nlk, Gurau:2017qya, Benedetti:2017qxl, Carrozza:2018ewt}. 

In view of these new developments, it is natural to wonder whether it is possible to construct SYK-like tensor models with a single flavour. In this paper we provide three examples, based on the three irreducible rank-$3$ representations of the \emph{compact symplectic group} $\Sp(N)$. The motivation for working with the symplectic group is that, unlike $\mO(N)$ \cite{Klebanov:2016xxf, Carrozza:2018ewt, TASI_tensor-review}, it allows to write a non-zero tetrahedral interaction for fermions in $d=1$. The price to pay is that one needs to work with complex fermions, which leads to a family of tensor model analogues of the complex SYK model \cite{Sachdev:1992fk, Sachdev:2015efa}. We emphasize that tensor models with symplectic symmetries have been previously considered \cite{Gubser:2017qed, Gubser:2018yec}, but in contrast to the present work, the focus was on Majorana fermions transforming under the \emph{non-compact} symplectic group $\Sp(2N, \mathbb{R})$.

Finally, we stress that the large $N$ structure of symplectic tensor models that we explicitly describe in this paper is not tied to the particular $d=1$ fermionic theory we choose to focus on, and could therefore be taken advantage of in other contexts. 

\medskip

The paper is organized as follows. We introduce the models and their symmetries in section \ref{sec:model}. In section \ref{sec:largeN} we describe their large $N$ and strong-coupling features. We then move on to the enumeration of singlet states by means of $\Sp(N)$ character integrals (section \ref{sec:singlets}), and we finally conclude our study by an explicit diagonalization of the two simplest instances of our models (section \ref{sec:diago}). Conventions as well as various technical details are relegated to the Appendix, which will be referred to whenever necessary.  

\section{Definition of the models}\label{sec:model}

\subsection{Fock space, Hamiltonian and action}

Let us consider a tensorial fermionic algebra of operators of the form:
\beq\label{algebra}
\{ \Gamma_{abc} , \Gamma_{a'b'c'} \} = 0 = \{ \Gamma_{abc}^\dagger , \Gamma_{a'b'c'}^\dagger \} \,, \qquad \{ \Gamma_{abc} , \Gamma_{a'b'c'}^\dagger \} = \bP_{abc , a'b'c'}\,,
\eeq 
where the tensor indices $a,b,c \ldots$ take value in $\{1, \ldots , 2N \}$ and $\bP$ is some symmetric kernel. We furthermore assume that $\Gamma^\dagger_{abc}$ transforms as a tensor product of three fundamental representations of $\Sp(N) = \U(2N) \cap \Sp(2N, \mathbb{C})$, namely\footnote{Summation over repeated indices is assumed throughout the paper, unless specified otherwise.}:
\beq\label{eq:action_spn}
\forall U \in \Sp(N)\,, \qquad [U \cdot \Gamma]_{abc} := U_{aa'}  U_{bb'}  U_{cc'} \Gamma_{a' b' c'} \,.
\eeq
We ensure that this action extends to an automorphism of the fermionic algebra by assuming $\bP$ to be the orthogonal projector associated to some $\Sp(N)$ rank-$3$ tensor representation. In order to guarantee the existence of a melonic large $N$ limit, it is furthermore crucial to eliminate all vector modes from this representation. This leaves us with only three inequivalent choices of irreducible representation:
\begin{enumerate}
\item $\bP = \bP^{(S)}$ is the orthogonal projector onto completely \emph{symmetric} tensors; 

\item $\bP = \bP^{(A)}$ is the orthogonal projector onto completely \emph{antisymmetric traceless} tensors; 

\item $\bP = \bP^{(M)}$ is the orthogonal projector onto \emph{mixed\footnote{By \emph{mixed symmetry tensor} we mean any tensor transforming under the two-dimensional irreducible representation of $\cS_3$ associated to the Young diagram $\mixednn$.} traceless} tensors. 
\end{enumerate}
A detailed construction of these representations, together with explicit expressions for $\bP^{(S)}$, $\bP^{(A)}$ and $\bP^{(M)}$, are provided in the Appendix \ref{app:projectors}. In the remainder of the paper, $\bP$ will denote any one of these three projectors; we will reserve the use of superscripts for investigations of specific features of the models $(S)$, $(A)$ and $(M)$. We also denote by $V$ the image of $\bP$ in the vector space of complex rank-$3$ tensors. $V$ has (complex) dimension $n:= \Tr \, \bP$, with: 
\begin{align}
n^{(S)} := \Tr \, \bP^{(S)} &= \frac{2N}{3} \left( 2 N^2 + 3 N + 1 \right) \,, \\
n^{(A)} := \Tr \, \bP^{(A)} &= \frac{2N}{3} \left( 2 N^2 - 3 N -2 \right) \,, \\
n^{(M)} := \Tr \, \bP^{(S)} &= \frac{8N}{3} \left( N^2 - 1 \right) \,.
\end{align}

\medskip

The Fock space generated by the algebra \eqref{algebra} is $\cF = \bigwedge(V) = \bigoplus_{k=0}^n \bigwedge^k (V)$, of dimension $d_{\cF} = 2^{n}$. In $\cF$ we consider the two-body Hamiltonian:
\beq\label{eq:hamiltonian}
H = \frac{g}{2} \epsilon_{bg} \epsilon_{dh} \Gamma_{abc}^\dagger \Gamma_{fge}^\dagger \Gamma_{ade} \Gamma_{fhc}\,,
\eeq
where $\epsilon$ is the invertible skew-symmetric matrix entering the definition of $\Sp(N)$ (which one may take to be \eqref{matrixJ}), and $g$ is a real coupling constant. $H$ is invariant under $\Sp(N)$, as an immediate consequence of the fact that:
\begin{itemize}
\item pairs of indices $a$ and $b$ belonging to operators of different types (i.e. one creation operator and one annihilation operator) are contracted with the $\U(2N)$ invariant bilinear $\delta_{ab}$;
\item pairs of contracted indices belonging to operators of the same type are contracted with the $\Sp(N,\mathbb{C})$ invariant bilinear $\epsilon_{ab}$.
\end{itemize} 

\medskip

In the path-integral formulation, the dynamics is encoded in time-dependent Grassmann variables $\{ \psib_{abc}(t), \psi_{abc}(t) \}$ governed by the action:
\beq\label{eq:action}
S[\psib, \psi] = \int \extd t \, \left( i 
\psib_{abc} \partial_t \psi_{abc}  - \frac{g}{2} \epsilon_{bg} \epsilon_{dh} \psib_{abc} \psib_{fge} \psi_{ade} \psi_{fhc} \right) \,.
\eeq
Importantly, the tensors $\psi_{abc}$ and $\psib_{abc}$ must be confined to the irreducible vector space associated to $\bP$; the total number of field components being summed over in the path-integral is therefore $2 \, \Tr \bP = 2 n$. Equivalently, one can view $\psi_{abc}$ and $\psib_{abc}$ as generic tensor fields, with bare propagator given by:
\beq
\langle T ( \psib_{abc}(t) \psi_{a' b' c'}(t') )\rangle_0 = G_0 (t-t') \, \bP_{abc, a' b' c'} = \mathrm{sgn}(t-t') \, \bP_{abc, a' b' c'}\,,
\eeq
which automatically projects the degrees of freedom down to the appropriate subspace.

The kernel of the $\Sp(N)$ invariant interaction term will be denoted $\cV_{{\mathbf{a}},{\mathbf{b}},{\mathbf{c}},{\mathbf{d}}}$ with the shorthand notation ${\mathbf{a}} = a_1 a_2 a_3$, ${\mathbf{b}} = b_1 b_2 b_3$ etc. Since the index contractions follow the combinatorial pattern of a tetrahedron, we will follow the literature and refer to it as a \emph{tetrahedral interaction}. We will furthermore use the following two graphical representations
\begin{align}
g \, \psib_{\mathbf{a}} \psi_{\mathbf{b}} \psib_{\mathbf{c}} \psi_{\mathbf{d}} \, \cV_{{\mathbf{a}},{\mathbf{b}},{\mathbf{c}},{\mathbf{d}}} &= g \, \psib_{\mathbf{a}} \psi_{\mathbf{b}} \psib_{\mathbf{c}} \psi_{\mathbf{d}} \, \epsilon_{a_2 c_2} \epsilon_{b_2 d_2} \delta_{a_1 b_1} \delta_{b_3 c_3} \delta_{c_1 d_1} \delta_{d_3 a_3}  \\
&= \psib_{\mathbf{a}} \psi_{\mathbf{b}} \psib_{\mathbf{c}} \psi_{\mathbf{d}} \quad \vcenter{\hbox{\includegraphics[scale=.9]{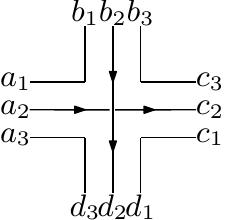}}} \label{vertex_strand}
\\
&= \, \vcenter{\hbox{\includegraphics[scale=.8]{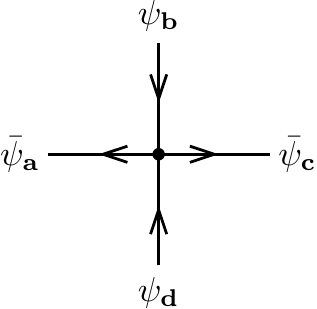}}} \label{vertex_arrow}
\end{align}
where the arrows in the first representation keep track of the ordering of indices in $\epsilon$ contractions, while the arrows in the second encode the type of field (by convention, $\psi$'s are attached to ingoing arrows, and $\psib$'s to outgoing ones). 

The kernel $\cV$ has an obvious symmetry under rotation by a $\pi$ angle: 
\beq
\cV_{{\mathbf{a}},{\mathbf{b}},{\mathbf{c}},{\mathbf{d}}} = \cV_{{\mathbf{c}},{\mathbf{d}},{\mathbf{a}},{\mathbf{b}}}\,. 
\eeq
Since the associated permutation is even, the interaction itself is invariant. In addition, the latter changes sign under a reflection about a vertical (or horizontal) axis:
\begin{align}
\psib_{\mathbf{a}} \psi_{\mathbf{b}} \psib_{\mathbf{c}} \psi_{\mathbf{d}} \, \cV_{{\mathbf{a}},{\mathbf{b}},{\mathbf{c}},{\mathbf{d}}} &= g \,\psib_{\mathbf{a}} \psi_{\mathbf{b}} \psib_{\mathbf{c}} \psi_{\mathbf{d}} \, \epsilon_{a_2 c_2} \epsilon_{b_2 d_2} \delta_{a_1 b_1} \delta_{b_3 c_3} \delta_{c_1 d_1} \delta_{d_3 a_3} \\
&= - g \,\psib_{\mathbf{a}} \psi_{\mathbf{b}} \psib_{\mathbf{c}} \psi_{\mathbf{d}} \, \epsilon_{c_2 a_2} \epsilon_{b_2 d_2} \delta_{a_1 b_1} \delta_{b_3 c_3} \delta_{c_1 d_1} \delta_{d_3 a_3} \\
&\underset{1 \leftrightarrow 3}{=}  - g \,\psib_{\mathbf{a}} \psi_{\mathbf{b}} \psib_{\mathbf{c}} \psi_{\mathbf{d}} \, \epsilon_{c_2 a_2} \epsilon_{b_2 d_2} \delta_{c_1 b_1} \delta_{b_3 a_3}  \delta_{d_1 a_1} \delta_{d_3 c_3}  \\
&= - \psib_{\mathbf{a}} \psi_{\mathbf{b}} \psib_{\mathbf{c}} \psi_{\mathbf{d}} \, \cV_{{\mathbf{c}},{\mathbf{b}},{\mathbf{a}},{\mathbf{d}}}\,,
\end{align}
where we have used that $\psi_{a_1 a_2 a_3} = \psi_{a_3 a_2 a_1}$ for the $S$ representation, and $\psi_{a_1 a_2 a_3} = - \psi_{a_3 a_2 a_1}$ for the $A$ and $M$ representations. Note at this stage how important it is to work with fermions: the bosonic theory with complex tensors $\phi_{abc}$ and $\bar\phi_{abc}$ (in any of the representations $S$, $A$ or $M$) has a vanishing tetrahedral interaction:
\beq
\bar\phi_{\mathbf{a}} \phi_{\mathbf{b}} \bar\phi_{\mathbf{c}} \phi_{\mathbf{d}} \, \cV_{{\mathbf{a}},{\mathbf{b}},{\mathbf{c}},{\mathbf{d}}} = - \bar\phi_{\mathbf{a}} \phi_{\mathbf{b}} \bar\phi_{\mathbf{c}} \phi_{\mathbf{d}} \, \cV_{{\mathbf{c}},{\mathbf{b}},{\mathbf{a}},{\mathbf{d}}} = - \bar\phi_{\mathbf{c}} \phi_{\mathbf{b}} \bar\phi_{\mathbf{a}} \phi_{\mathbf{d}} \, \cV_{{\mathbf{c}},{\mathbf{b}},{\mathbf{a}},{\mathbf{d}}} = 0\,. 
\eeq
Instead, in our fermionic model the statistics and the antisymmetry of $\epsilon$ conspire to make the interaction non-trivial. In fact, comparing with the tensor quantum mechanics of $\mO(N)$ irreducible tensors, we find that the situation is precisely reversed: if the two $\epsilon$'s in $\cV$ are replaced by $\delta$'s, a similar argument lead to opposite signs and as a result the interaction vanishes for fermions \cite{Klebanov:2016xxf, Carrozza:2018ewt, TASI_tensor-review}. These elementary considerations are summarized in Table \ref{table-spnon}.

\begin{table}[h]
\centering
\begin{tabular}{|c|c|c|}\hline
& $\mO(N)$ irreducible tensors & $\Sp(N)$ irreducible tensors \\\hline
Bosonic statstics & $\neq0$ & $=0$ \\\hline
Fermionic statistics & $=0$ & $\neq0$ \\\hline
\end{tabular}
\caption{Nature of the tetrahedral interaction as a function of the statistics.}\label{table-spnon}
\end{table}

\medskip

We find an additional subtlety in the mixed representation $M$: because of the two-dimensional nature of the $\cS_3$ representation associated to the Young tableau $\mixed$, the tetrahedral interaction is not unique. The space of tetrahedral interactions turns out to be itself two-dimensional, and the most general action we will consider in this case is:
\begin{align}\label{vertex_mixed}
S_{\mathrm{int}}^{(M)} &= \frac{g_1}{2} \epsilon_{bg} \epsilon_{dh} \psib_{abc} \psib_{fge} \psi_{ade} \psi_{fhc} + \frac{g_2}{2} \epsilon_{ad} \epsilon_{ef} \psib_{abc} \psib_{ebg} \psi_{dhg} \psi_{fhc} \\
&= \psib_{\mathbf{a}} \psi_{\mathbf{b}} \psib_{\mathbf{c}} \psi_{\mathbf{d}}  \left( \frac{g_1}{2} \quad \vcenter{\hbox{\includegraphics[scale=.9]{VertexBlack2}}} + \frac{g_2}{2} \quad \vcenter{\hbox{\includegraphics[scale=.9]{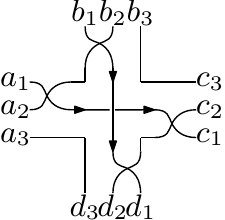}}} \right)\,,
\end{align}
where the second term is simply obtained by the action of the permutation $(12)$. The fact that no other independent term can be constructed is a consequence of the identity
\beq
[(12)+(23)] \triangleright T = T
\eeq 
in the representation $\mixed$. 

\medskip 

Finally, we note that the generalized vertex \eqref{vertex_mixed} has the same rotational symmetry as $\cV$ under rotations by $\pi$. On the other hand it is not symmetric under reflections: in contrast to the $S$ and $A$ representations, the field does not transform by a simple sign under the permutation $(12)$, and as a consequence the term proportional to $g_2$ in \eqref{vertex_mixed} is not invariant. This means that, strictly speaking, the amplitudes of this model cannot unambiguously be labelled by graphs. One must instead carefully keep track of the local embedding of the interaction vertices. It is convenient in such situation to use the language of \emph{embedded graphs}, also known as \emph{combinatorial maps}, and to speak of Feynman maps rather than Feynman graphs. We refer to \cite{Carrozza:2018ewt} for more detail in the similar context of mixed traceless $\mO(N)$ tensor models. For clarity of the exposition and given that this subtlety only affects the representation $M$, we will largely ignore this point and pretend that the Feynman amplitudes can be unambiguously labelled by $4$-regular diagrams with oriented lines and vertices as in \eqref{vertex_arrow}.

\subsection{Symmetries and charges}

The action \eqref{eq:action} is invariant under the global symmetry group $\Sp(N) \times \U(1)$. The $\U(1)$ symmetry $\psi \mapsto e^{i \theta} \psi$ implies the conservation of the fermionic number operator:
\beq
\cQ := \frac{1}{2} [ \psib_{abc} , \psi_{abc} ]\,.
\eeq

\medskip

As for the $N(2N+1)$ charges associated to the $\Sp(N)$ symmetry \eqref{eq:action_spn}, they can be inferred from the action of the generators constructed in the Appendix \eqref{ap:generators}. One obtains: 
\begin{align}
\hat{I}_{k,l} &: = \left[ i \left( \psib_{(2k-1)bc} \psi_{(2l-1)bc} + \psib_{(2k)bc} \psi_{(2l)bc} \right)  \; + \; \mathrm{c.c} \; \right] \quad + \quad (1 \rightarrow 2 \rightarrow 3) \,, & \nn \\
\hat{\Sigma}^{(1)}_{k,l} &:= \left[  \left( \psib_{(2k-1)bc} \psi_{(2l)bc} + \psib_{(2k)bc} \psi_{(2l-1)bc} \right)  \; + \; \mathrm{c.c} \; \right] \quad + \quad (1 \rightarrow 2 \rightarrow 3) \,, & \nn \\
\hat{\Sigma}^{(2)}_{k,l} &:= \left[  i \left( \psib_{(2k-1)bc} \psi_{(2l)bc} - \psib_{(2k)bc} \psi_{(2l-1)bc} \right)  \; + \; \mathrm{c.c} \; \right] \quad + \quad (1 \rightarrow 2 \rightarrow 3) \,, & \nn \\
\hat{\Sigma}^{(3)}_{k,l} &:= \left[ \left( \psib_{(2k-1)bc} \psi_{(2l-1)bc} - \psib_{(2k)bc} \psi_{(2l)bc} \right)  \; + \; \mathrm{c.c} \; \right] \quad + \quad (1 \rightarrow 2 \rightarrow 3) \,, &\; 1 \leq k < l \leq N\,, \nn \\
\hat{\Sigma}^{(1)}_m &:= \left( \psib_{(2m-1)bc} \psi_{(2m)bc} + \psib_{(2m)bc} \psi_{(2m-1)bc} \right)   \quad + \quad (1 \rightarrow 2 \rightarrow 3) \,, & \nn \\
\hat{\Sigma}^{(2)}_m &:= i \left( \psib_{(2m-1)bc} \psi_{(2m)bc} - \psib_{(2m)bc} \psi_{(2m-1)bc} \right)   \quad + \quad (1 \rightarrow 2 \rightarrow 3) \,, & \nn \\
\hat{\Sigma}^{(3)}_m &:= \left( \psib_{(2m-1)bc} \psi_{(2m-1)bc} - \psib_{(2m)bc} \psi_{(2m)bc} \right)   \quad + \quad (1 \rightarrow 2 \rightarrow 3)  \,, &\; 1 \leq m \leq N\,, \nn
\end{align}
where $(1 \rightarrow 2 \rightarrow 3)$ indicates a sum of two terms obtained by cyclic permutation of the indices\footnote{Those three terms are identical in the representations $A$ and $S$, but not necessarily in the representation $M$.}. 
The quadratic Casimir operator is then (proportional to)
\beq\label{eq:casimir}
\cC_2 := \sum_{1 \leq k<l\leq N} \left( ( \hat{I}_{k,l})^2 + \sum_{p =1}^3 (\hat{\Sigma}^{(p)}_{k,l})^2 \right) + 2 \sum_{1 \leq m \leq N} \sum_{p = 1}^3 (\hat{\Sigma}^{(p)}_m)^2 \,. 
\eeq 

In holographic applications of tensor models, the states of interest are singlets, a restriction to which can be enforced by a gauging procedure \cite{Klebanov:2016xxf}. We will ignore this point in most of the text, except in section \ref{sec:singlets} when we will enumerate those singlets.

\section{Large $N$ regime}\label{sec:largeN}

\subsection{Existence of the large $N$ expansion}

The models \eqref{eq:action} and \eqref{vertex_mixed} admit a large $N$ expansion with respect to the 't Hooft coupling:
\beq
\lambda = N^{3/2} g \qquad \mathrm{(or} \; \lambda_i = N^{3/2} g_i \,, i=1,2  \mathrm{)}\,.
\eeq 
Its existence can be proven by means of a general method developed for (anti)-symmetric (traceless) $\mO(N)$ tensors in \cite{Benedetti:2017qxl}, and generalized to mixed traceless tensors in \cite{Carrozza:2018ewt}. We will not reproduce the full construction here, but only briefly explain why it applies equally well in the symplectic context, as long as one makes sure to work with an irreducible representation.
 
\begin{figure}[htb]
\centering
\includegraphics[scale=1]{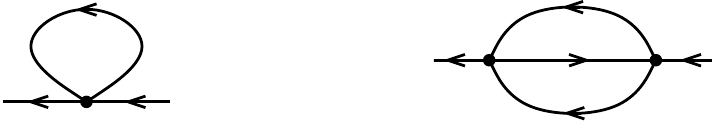}
\caption{The elementary tadpole (left) and the elementary melon (right).}\label{fig:melon-tadpole}
\end{figure}

Two classes of diagrams play a central role in this discussion: \emph{tadpoles} and \emph{melons}. The elementary tadpole and melon diagrams shown in Figure \ref{fig:melon-tadpole} recursively generate the family of \emph{melon-tadpole} diagrams, while the elementary melon alone generates the subclass of \emph{melon diagrams}. A salient feature of tensor models in general is that they are typically dominated by melon diagrams in the large $N$ limit \cite{critical, RTM}.
 
Each Feynman diagram can be decomposed into a sum of \emph{stranded graphs} resulting from the representations \eqref{vertex_strand} (or \eqref{vertex_mixed}) of the vertex, and the representations \eqref{projS}, \eqref{projA} or \eqref{projM} of the projector $\bP$. Just like in $\mO(N)$ models \cite{Benedetti:2017qxl}, a $2$-point stranded graph $G$ can be shown to scale as:
\beq\label{omega}
N^{-\omega(G)}\,, \qquad  \omega(G) := 3 + \frac{3}{2} V(G) + B(G) - F(G)\,,
\eeq 
where $V$ is the number of vertices, $B$ is the number of broken propagators (i.e. terms involving $\epsilon$ contractions in \eqref{projS}, \eqref{projA} or \eqref{projM}), and $F$ is the number of \emph{faces} (i.e. the number of closed cycles formed by the strands). The only difference is that, due to the antisymmetry of $\epsilon$, any stranded configuration containing a face involving an odd number of $\epsilon$ contractions vanishes.  

The proof itself can then proceed in two main steps:
\begin{enumerate}
\item it is shown that $\omega(G) > 0$ for any stranded configuration with neither \emph{tadpoles} nor \emph{melons}, thus demonstrating that such diagrams are always suppressed in the large $N$ limit;

\item it is then proven that melon-tadpoles can be resummed in a controlled manner, in such a way that only melon diagrams survive at leading order in $N$.
\end{enumerate} 
The first claim results from purely combinatorial considerations \cite{Benedetti:2017qxl} and automatically applies to $\Sp(N)$ since: a) the set of stranded configurations generated by $\Sp(N)$ models is strictly contained in the set of $\mO(N)$ stranded graphs; b) the $\Sp(N)$ and $\mO(N)$ power countings \eqref{omega} agree in these two sets. The second claim relies on precise cancellations between the certain stranded configurations of certain generalized tadpole diagrams, which would otherwise generate unbounded scalings. A nice feature of the argument is that such cancellations automatically occur in an irreducible tensor representation \cite{Carrozza:2018ewt}. 

\medskip

We conclude this section with the main combinatorial identities relevant to the computation of melonic amplitudes in the large $N$ limit. Ignoring the time dependence, the contraction of projectors associated to an elementary melon diagram is:
\beq
\vcenter{\hbox{\includegraphics[scale=.9]{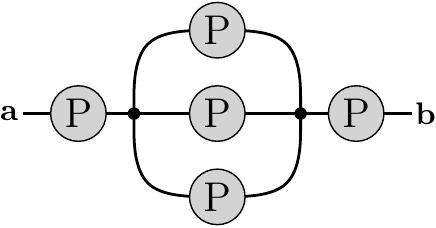}}} := \bP_{{\mathbf{a}}{\mathbf{c}}} \cV_{{\mathbf{c}} {\mathbf{d}} {\mathbf{e}} {\mathbf{f}}} \bP_{{\mathbf{d}}{\mathbf{i}}} \bP_{{\mathbf{e}}{\mathbf{h}}} \bP_{{\mathbf{f}}{\mathbf{g}}}    \cV_{{\mathbf{g}} {\mathbf{h}} {\mathbf{i}} {\mathbf{j}}} \bP_{{\mathbf{j}}{\mathbf{b}}} \,.
\eeq 
The irreducibility of the tensor representation implies -- by Schur's lemma -- that this contraction is proportional to $\bP$ itself:
\beq
\vcenter{\hbox{\includegraphics[scale=.9]{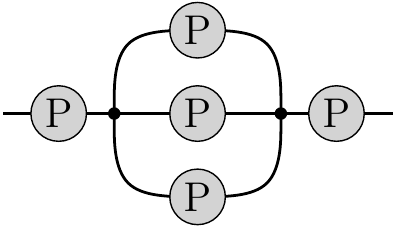}}} = \cM_N \, \vcenter{\hbox{\includegraphics[scale=.9]{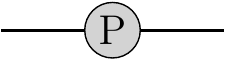}}}\,. 
\eeq
More explicitly, we can compute the coefficient of proportionality $\cM_N$ and check that it does indeed scale as $O(1)$ at large $N$. Exact expressions can be found in the Appendix \ref{app:Melons}. In the rest of the text, we will only make use of the limits: 
\begin{align}
\cM^{(S)}_N & \underset{N\to \infty}{\longrightarrow} \frac{\lambda^2}{27} =: \cM^{(S)}
\,,\\
\cM^{(A)}_N & \underset{N\to \infty}{\longrightarrow}  \frac{\lambda^2}{27} =: \cM^{(A)} 
\,,\\
\cM^{(M)}_N & \underset{N\to \infty}{\longrightarrow} \frac{8 {\lambda_1}^2 + \lambda_1 \lambda_2 + 8 {\lambda_2}^2}{27} =: \cM^{(M)}   
\,.
\end{align}
In the representation $M$, one may construct a second inequivalent melonic $2$-point map, namely: 
\beq
\vcenter{\hbox{\includegraphics[scale=.9]{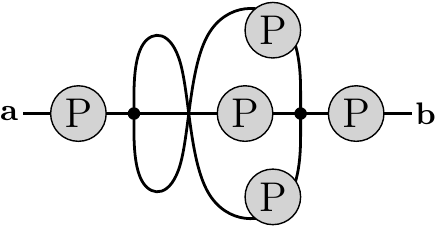}}} := \bP_{{\mathbf{a}}{\mathbf{c}}} \cV_{{\mathbf{c}} {\mathbf{d}} {\mathbf{e}} {\mathbf{f}}} \bP_{{\mathbf{d}}{\mathbf{g}}} \bP_{{\mathbf{e}}{\mathbf{h}}} \bP_{{\mathbf{f}}{\mathbf{i}}}    \cV_{{\mathbf{g}} {\mathbf{h}} {\mathbf{i}} {\mathbf{j}}} \bP_{{\mathbf{j}}{\mathbf{b}}} = - \cN^{(M)}_N \, \bP_{\mathbf{a}\mathbf{b}} \,, 
\eeq
and we compute
\beq
\cN^{(M)}_N  \underset{N\to \infty}{\longrightarrow} \frac{16 {\lambda_1}^2 + 2 \lambda_1 \lambda_2 +  {\lambda_2}^2}{54} =: \cN^{(M)}\,.
\eeq
For later use, we finally define the effective infrared coupling constant:
\begin{equation}
\lambda_\mathrm{eff}^2 := \left\{ 
\begin{aligned}
&2 \cM^{(r)} = \frac{2 \lambda^2}{27}  &\qquad {\mathrm{for}}\; r=S \;{\mathrm{or}}\; A \\
&\cM^{(r)} + \cN^{(r)} = \frac{32 {\lambda_1}^2 + 4 \lambda_1 \lambda_2 + 17 {\lambda_2}^2}{54}  &\qquad {\mathrm{for}} \; r=M 
\end{aligned}\right.
\end{equation}

\subsection{Two-point function in the infrared limit}

As is now standard, the two-point function of a melonic quantum mechanics typically develops an emergent conformal invariance at strong coupling\footnote{By which we mean the regime $\vert \lambda_\mathrm{eff}^2 / \omega \vert \gg 1$ where $\omega$ is the Fourier conjugate of $t$. It can therefore be equivalently understood as an infrared limit.}. Being complex, our model is more closely related to the complex SYK model \cite{Sachdev:2015efa}, or to the $\SU(N)\times \mO(N) \times \SU(N)$ tensor quantum mechanics reviewed in \cite{Klebanov:2016xxf, TASI_tensor-review}.

By virtue of the $\Sp(N)$ invariance and the irreduciblity of the chosen tensor representation, the full two-point function must take the form:
\beq
\langle T ( \psib_{abc}(t) \psi_{a' b' c'}(t') )\rangle = G (t,t') \, \bP_{abc, a' b' c'} = G (t-t') \, \bP_{abc, a' b' c'} \,,
\eeq
where $G(-t) = -G(t)$. In the large $N$ limit, it furthermore reduces to an infinite sum of melon diagrams. Given the simple recursive structure of the latter, this leads to the closed Schwinger-Dyson equation:
\beq
G = G_0 + \le\, G_0 \ast G^3 \ast G\,,
\eeq 
where $\ast$ is the convolution product $[f\ast g] (t_1 , t_3) := \int \extd t_2 \,f (t_1,t_2) g (t_2 , t_3) $. As one subsequently takes the infrared limit, only the right-hand side survives and one finally obtains:
\beq\label{SDE-IR}
G \ast G^3 = \frac{-1}{\le}\,.
\eeq
The remarkable feature of this equation is that it is (formally) invariant under diffeomorphisms $t \mapsto f(t)$ (which is of course the root cause of the solvable nature of SYK-like theories):
\beq
G(t_1 ,t_2) \mapsto \vert f'(t_1 ) f'(t_2 )\vert^{1/4} G(f(t_1) , f(t_2))\,.
\eeq
The (time-translation invariant) solution of \eqref{SDE-IR} is then conformal with dimension $\Delta=1/4$ \cite{Klebanov:2016xxf, Gurau:2017qna}:
\beq\label{sol2point}
G(t_1 ,t_2 ) = - \left( \frac{1}{4 \pi \le}\right)^{1/4} \frac{\mathrm{sgn}(t_1-t_2)}{\vert t_1-t_2 \vert^{1/2}} \,. 
\eeq

\subsection{Four-point function and conformal spectrum}

\begin{figure}[htb]
\centering
\includegraphics[scale=.9]{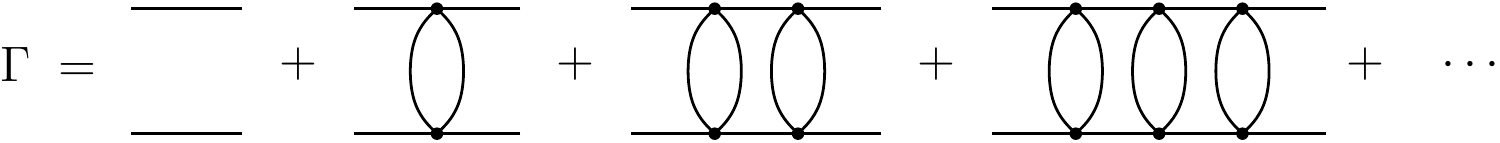}

\vspace{.5cm}

\includegraphics[scale=.9]{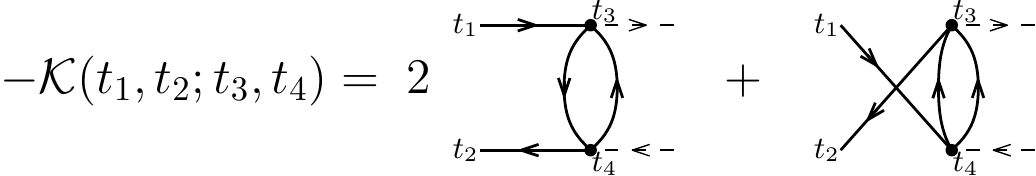}
\caption{Top: the sum over ladder diagrams computed by $\Gamma$. Bottom: structure of the rung operator $\cK$. Lines represent full propagators, and arrows are kept implicit in the top picture.}\label{fig:ladder-sum}
\end{figure}

The conformal spectrum of bilinear operators in the infrared regime can be extracted from the $4$-point correlator, which at large $N$ decomposes as:
\beq
\langle \psib_{abc} (t_1) \psi_{abc} (t_2) \psib_{def} (t_3) \psi_{def} (t_4)  \rangle = n^2 G(t_1,t_2) G(t_3,t_4) + n \Gamma(t_1 , t_2 , t_3, t_4) + O(N^{5/2}) \,,
\eeq 
where we remind the reader that $n = \Tr\,\bP \sim N^3$. The second term is a sum of ladder diagrams as illustrated in the top half of Figure \ref{fig:ladder-sum}. As a result, it has the structure
\beq
\Gamma = \sum_{p = 0}^\infty \, \Gamma_p
\eeq
with:
\begin{align}
\Gamma_0 ( t_1 , t_2 , t_3 , t_4 ) &= G(t_1, t_4) G(t_2, t_3)\,, \\
\Gamma_{p+1}( t_1 , t_2 , t_3 , t_4 )  &= \int \extd t \extd t' \, \cK  ( t_1 , t_2 ; t , t') \Gamma_{p}( t , t' , t_3 , t_4 )\,,
\end{align} 
and where $\cK$ is the kernel associated to the operator that adds a rung to a ladder diagram. In the present model there are two ways of doing so, as represented in the bottom half of Figure \ref{fig:ladder-sum}, which yields
\begin{align}
\cK( t_1 , t_2 ; t_3 , t_4 ) = - \le \left[2 G(t_1, t_3) G(t_2, t_4) - G(t_1, t_4) G(t_2, t_3) \right] G(t_3, t_4)^2 \,.
\end{align}
The combinatorial factor $2$ in the first term is a consequence of the opposite relative orientations of the two lines making up the rung, while the minus sign in the second is due to the fermionic statistics. 
Plugging the solution \eqref{sol2point} in, the dependence in the effective coupling constant $\le$ drops out and we obtain:
\beq\label{ladder-op}
\cK( t_1 , t_2 ; t_3 , t_4 ) =  \frac{-1}{4 \pi \vert t_3 - t_4 \vert} \left( \frac{2 \mathrm{sgn}(t_1 - t_3) \mathrm{sgn}(t_2 - t_4)}{\vert t_1 - t_3 \vert^{1/2} \vert t_2 - t_4 \vert^{1/2}} - \frac{\mathrm{sgn}(t_1 - t_4) \mathrm{sgn}(t_2 - t_3)}{\vert t_1 - t_4 \vert^{1/2} \vert t_2 - t_3 \vert^{1/2}} \right)\,.
\eeq

\begin{figure}[htb]
\centering
\includegraphics[scale=.7]{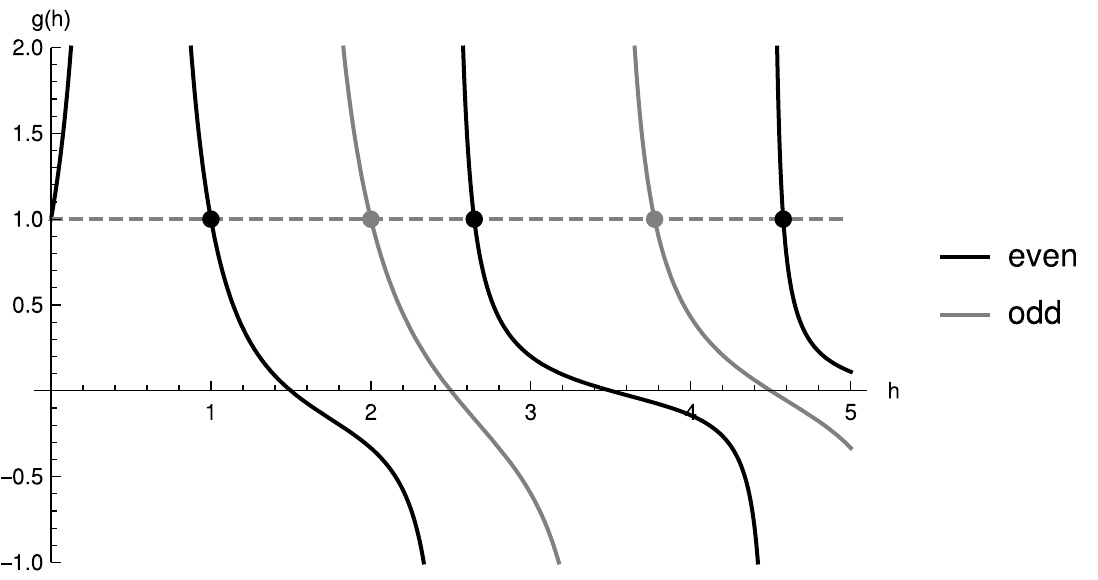}
\caption{Conformal dimensions from the eigenvalue equations $g_{\mathrm{even}}(h)=1$ and $g_{\mathrm{odd}}(h)=1$.}\label{fig:conformal-plot}
\end{figure}

The rung operator \eqref{ladder-op} is identical to that found in the $\SU(N) \times \mO(N) \times \SU(N)$ tensor quantum mechanics discussed in \cite{Klebanov:2016xxf, TASI_tensor-review}. As a result we find the same conformal spectrum as derived there, for all bilinear primary operators of the form:
\beq
\cO_2^k = \psib_{abc} ( \partial_t^k \psi)_{abc} + \cdots \,, \qquad k \in \mathbb{N}\,.
\eeq
Let us briefly summarize those findings. Following \cite{Maldacena:2016hyu, Polchinski:2016xgd, Gross:2016kjj}, the conformal dimensions $h_k$ can be determined from an analogue of the Bethe-Salpeter equation. It consists in the eigenvalue equation
\beq
v_k (t_0 , t_1 , t_2) = g_k (h_k) \int \extd t_3 \extd t_4 \cK(t_1 , t_2 ; t_3, t_4) v_k (t_0 , t_3 , t_4) \quad \mathrm{with} \quad g_k (h_k) = 1
\eeq
for the three-point correlator
\beq
v_k (t_0 , t_1 , t_2) := \langle \cO_2^k (t_0) \psi_{abc} (t_1) \psib_{abc} (t_2) \rangle = \left\lbrace \begin{aligned} \frac{c_k \mathrm{sgn}( t_0 - t_1 ) \mathrm{sgn}( t_0 - t_2 )}{\vert t_0 - t_1 \vert^{h_k} \vert t_0 - t_2 \vert^{h_k} \vert t_1 - t_2 \vert^{1/2-h_k}} &\quad \mathrm{if}\;k \; \mathrm{is}\; \mathrm{even} \\
\frac{c_k \mathrm{sgn}( t_1 - t_2 )}{\vert t_0 - t_1 \vert^{h_k} \vert t_0 - t_2 \vert^{h_k} \vert t_1 - t_2 \vert^{1/2-h_k}} &\quad \mathrm{if}\;k \; \mathrm{is}\; \mathrm{odd}
\end{aligned} \right. 
\eeq 
This self-consistency equation is independent of the OPE coefficient $c_k$, and since we also know the rung operator \eqref{ladder-op}, the function $g_k$ may be explicitly computed. One finds \cite{TASI_tensor-review}:
\beq
g_k (h) = \left\lbrace \begin{aligned} g_{\mathrm{even}} (h) &:= - \frac{\tan(\frac{\pi}{2}(h+\frac{1}{2}))}{2 h - 1} &\quad \mathrm{if}\;k \; \mathrm{is}\; \mathrm{even} \\
g_{\mathrm{odd}} (h) &:= - \frac{3 \tan(\frac{\pi}{2}(h-\frac{1}{2}))}{2 h - 1} &\mathrm{if}\;k \; \mathrm{is}\; \mathrm{odd}
\end{aligned} \right. 
\eeq
The solutions of $g_{\mathrm{even}} (h) =1$ and $g_{\mathrm{odd}}(h) = 1$ are the conformal dimensions:
\beq
h_0 = 1 \,, \quad h_1 = 2 \,, \quad h_2 \approx 2.65 \,, \qquad h_3 \approx 3.77 \,, \qquad \mathrm{etc.}
\eeq
See Figure \ref{fig:conformal-plot}. The two exact solutions $h_0$ and $h_1$ are modes respectively associated to the $\U(1)$ and diffeomorphism symmetries \cite{Sachdev:2015efa, Maldacena:2016hyu, Kitaev:2017awl}. 

\medskip

Thanks to the 2PI formalism introduced in \cite{Benedetti:2018goh}, an effective action for tensor models that reproduces the key features of the bilocal collective field theory of the SYK model can be devised. As anticipated in \cite{Choudhury:2017tax, Bulycheva:2017ilt} and confirmed in \cite{Benedetti:2018goh}, tensor models turn out to have an important extra feature: in addition to the pseudo-Goldstone modes already present in SYK (the $h_0$ and $h_1$ modes), the global symmetry of tensor models is responsible for the presence of a large number of extra pseudo-Goldstone modes. Specifying to our model, we find $N(2N+1)$ zero modes associated to the enhancement of the global $\Sp(N)$ symmetry into a gauge symmetry in the infinite coupling limit. The effective dynamics of these infrared modes can then be determined by evaluating the leading-order inverse coupling correction, which breaks the emergent local symmetry. Following \cite{Choudhury:2017tax, Benedetti:2018goh, Maldacena:2016hyu}, the resulting effective field theory is a $\Sp(N)$ non-linear sigma model. 

\section{Enumeration of singlet states}\label{sec:singlets}

In this section we count the number of singlets by means of standard group-theoretic techniques. In particular, we follow the same approach as in \cite{Aharony:2003sx, Beccaria:2017aqc, Klebanov:2018nfp}. Other combinatorial methods for enumerating tensor invariants can be found in \cite{Geloun:2013kta, deMelloKoch:2017bvv, BenGeloun:2017vwn}.

\subsection{Generating function of singlet states}

The number of $\Sp(N)$ singlets in the Fock space $\cF$ may be inferred from character integrals. Let us denote by $\chi = \chi_\rho$ the character of the irreducible representation $\rho$ associated to $\bP$. From \eqref{projS}, \eqref{projA} and \eqref{projM}, one finds:
\begin{align}
\chi^{(S)} (U) &= \frac{1}{6} \tr( U )^3 + \frac{1}{2} \tr( U^2 ) \tr(U) + \frac{1}{3} \tr( U^3 ) \,, \label{chiS} \\
\chi^{(A)} (U) &= \frac{1}{6} \tr( U )^3 - \frac{1}{2} \tr( U^2 ) \tr(U) + \frac{1}{3} \tr( U^3 ) - \tr(U) \,, \label{chiA} \\
\chi^{(M)} (U)&= \frac{1}{3} \tr( U )^3 - \frac{1}{3} \tr( U^3 ) - \tr(U) \label{chiM} \,.
\end{align}
The character of the representation $\wedge(\rho)$ induced by $\rho$ in $\cF$ is:
\beq
\chi_{\wedge(\rho)} (U) = \sum_{k=0}^n \chi_{\wedge^k(\rho)} = \sum_{k=0}^n \Tr[ \wedge^k (\rho(U)) ] =  \det \left[ 1 + \rho(U) \right]
\eeq
Finally, the number $I_N$ of $\Sp(N)$ singlets in $\cF$ is the integral of $\chi_{\wedge(\rho)}$ with respect to the Haar measure, namely:
\beq
I_N 
= \int_{\Sp(N)} \extd U \, \det \left[ 1 + \rho(U) \right] 
\eeq
The integrand being a class function, this expression reduces to a $N$-dimensional integral of the form \eqref{int_class}, and can be evaluated by numerical methods when $N$ is not too large. 

More generally, the generating function of singlet operators is:
\beq\label{I_t}
I_N (t) := \int_{\Sp(N)} \extd U \, \det \left[ 1 + t \rho(U) \right] =  \int_{\Sp(N)} \extd U \, \exp\left( - \sum_{k = 1}^{+\infty} \frac{(-t)^k}{k} \chi( U^k ) \right) \,. 
\eeq

\subsection{Total number of singlets at small $N$}

\subsubsection{Symmetric sector}

Given $U \in \Sp(N)$ with eigenvalues $\{ e^{\pm i \theta_k}, \, k = 1, \ldots , N\}$, let us define 
\beq
\forall 1 \leq k \leq N\,, \qquad \tilde\theta_{2k-1} = \theta_k \qquad \mathrm{and} \qquad  \tilde\theta_{2k} = - \theta_k \,. 
\eeq
The matrix $\rho^{(S)}(U)$ has then eigenvalues $\{ e^{i ( \tilde\theta_k + \tilde\theta_l + \tilde\theta_m )}, \, 1 \leq k \leq l \leq m \leq 2N \}$, and hence:
\beq
\det \left[ 1 + \rho(U) \right] = \prod_{1 \leq k \leq l \leq m \leq 2N} \left( 1 + e^{i ( \tilde\theta_k + \tilde\theta_l + \tilde\theta_m )} \right) = 2^{n^{(S)}}  \prod_{1 \leq k \leq l \leq m \leq 2N} \cos \frac{ \tilde\theta_k + \tilde\theta_l + \tilde\theta_m }{ 2}\,.
\eeq
Henceforth
\beq\label{Isym_init}
I_N^{(S)} = 2^{\frac{2N}{3} \left( 2 N^2 + 3 N + 1 \right)} \int_{[- \pi , \pi]^N} \extd\mu( \theta_1 , \, \ldots , \, \theta_N) \,  \prod_{1 \leq k \leq l \leq m \leq 2N} \cos \frac{ \tilde\theta_k + \tilde\theta_l + \tilde\theta_m }{ 2}\,,
\eeq
where the measure $\extd \mu$ is given by \eqref{measure} in Appendix \ref{sec:int-formula}. After some algebra (see the Appendix \ref{sec:chi-formulas}), this formula can finally be reorganized as
\begin{align}\label{Isym_final}
I_N^{(S)} &= 2^{\frac{2N}{3} \left( 2 N^2 + 3 N + 1 \right)} \int_{[- \pi , \pi]^N} \extd\mu( \theta_1 , \, \ldots , \, \theta_N) \,  \left(\prod_{k = 1}^N \cos \frac{3 \theta_k}{2}\right)^2 \left( \prod_{k = 1}^N \cos\frac{\theta_k}{2} \right)^{2N} \nn \\
& \qquad \qquad \times \left( \prod_{1 \leq k < l \leq N } \cos \frac{2 \theta_k + \theta_l }{2} \cos \frac{ \theta_k + 2 \theta_l }{2} \cos \frac{2 \theta_k - \theta_l }{2} \cos \frac{ \theta_k - 2 \theta_l }{2} \right)^2 \\
& \qquad  \qquad \times \left( \prod_{1 \leq k < l < m \leq N} \cos \frac{\theta_k + \theta_l + \theta_m}{2} \cos \frac{\theta_k + \theta_l - \theta_m}{2} \cos \frac{\theta_k - \theta_l + \theta_m}{2} \cos \frac{\theta_k - \theta_l - \theta_m}{2} \right)^2 \nn
\end{align}

\medskip

Upon (numerical) integration we find:
\beq\label{enumS_r}
I_1^{(S)} = 3 \,, \qquad I_2^{(S)} = 39 \qquad \mathrm{and} \qquad  I_3^{(S)} = 170640\,.  
\eeq

\subsubsection{Antisymmetric traceless sector}

The eigenvalues of $\rho^{(A)}(U)$ are the elementary symmetric polynomials of eigenvalues of $U$ $$\{ e^{i ( \tilde\theta_k + \tilde\theta_l + \tilde\theta_m )}, \, 1 \leq k < l < m \leq 2N \}\,$$ minus the eigenvalues associated to the trace modes. The latter are $\{ e^{i \tilde\theta_k }, \, 1 \leq k \leq 2N\}$. Following the same procedure as in the symmetric case, we obtain
\beq\label{Iasym_init}
I_N^{(A)} = 2^{n^{(A)}} \int_{[- \pi , \pi]^N} \extd\mu( \theta_1 , \, \ldots , \, \theta_N) \,  \frac{\underset{1 \leq k < l < m \leq 2N}{\prod} \cos \frac{ \tilde\theta_k + \tilde\theta_l + \tilde\theta_m }{ 2}}{\underset{1\leq k \leq N}{\prod} \cos^2 \frac{\theta_k}{2}}\,,
\eeq
or more explicitly

\begin{align}\label{Iasym_final}
I_N^{(A)} &= 2^{\frac{2N}{3} \left( 2 N^2 - 3 N -2 \right)} \int_{[- \pi , \pi]^N} \extd\mu( \theta_1 , \, \ldots , \, \theta_N) \, \left( \prod_{k=1}^N \cos \frac{\theta_k}{2} \right)^{2N - 4} \\ 
& \qquad \qquad \times \left( \prod_{1 \leq k < l < m \leq N} \cos \frac{\theta_k + \theta_l + \theta_m}{2} \cos \frac{\theta_k + \theta_l - \theta_m}{2} \cos \frac{\theta_k - \theta_l + \theta_m}{2} \cos \frac{\theta_k - \theta_l - \theta_m}{2} \right)^2 \,. \nn
\end{align}

\medskip

We evaluated the first two non-zero values of $I_N^{(A)}$ numerically, and found:
\beq\label{enumA_r}
I_3^{(A)} = 8 \qquad \mathrm{and} \qquad I_4^{(A)} = 370 \,. 
\eeq

\subsubsection{Mixed traceless sector}

From \eqref{chiM} and \eqref{I_t}, one may directly infer that:
\beq\label{Im_init}
I_N^{(M)} = 2^{n^{(M)}} \int_{[- \pi , \pi]^N} \extd\mu( \theta_1 , \, \ldots , \, \theta_N) \,  \left( \frac{\underset{k , l , m = 1 ,\, \ldots ,\, N}{\prod} \cos \frac{ \tilde\theta_k + \tilde\theta_l + \tilde\theta_m }{ 2}}{\underset{1\leq k \leq N}{\prod} \cos^2 \frac{3\theta_k}{2}} \right)^{1/3} \times \frac{1}{\prod_{k = 1}^N \cos^2 \frac{3\theta_k}{2}}\,,
\eeq
which, by means of the elementary manipulations laid out in the Appendix \ref{sec:chi-formulas}, can be expressed in the form
\begin{align}\label{Im_final}
I_N^{(M)} &= 2^{\frac{8N}{3} \left( N^2 - 1 \right)} \int_{[- \pi , \pi]^N} \extd\mu( \theta_1 , \, \ldots , \, \theta_N) \,  \left( \prod_{k = 1}^N \cos\frac{\theta_k}{2} \right)^{4N-4} \nn \\
& \qquad \qquad \times \left( \prod_{1 \leq k < l \leq N } \cos \frac{2 \theta_k + \theta_l }{2} \cos \frac{ \theta_k + 2 \theta_l }{2} \cos \frac{2 \theta_k - \theta_l }{2} \cos \frac{ \theta_k - 2 \theta_l }{2} \right)^2 \\
& \qquad \qquad \times \left( \prod_{1 \leq k < l < m \leq N} \cos \frac{\theta_k + \theta_l + \theta_m}{2} \cos \frac{\theta_k + \theta_l - \theta_m}{2} \cos \frac{\theta_k - \theta_l + \theta_m}{2} \cos \frac{\theta_k - \theta_l - \theta_m}{2} \right)^4 \,. \nn
\end{align}

\medskip

The first non-zero values of $I_N^{(M)}$ are found to be
\beq
I_2^{(M)} = 18 \qquad \mathrm{and}  \qquad I_3^{(M)} = 169826605 \,.
\eeq 

\subsubsection{Summary}

We summarize our findings in Table \ref{tab:counting}. We notice that in all three representations, only a handful of singlets are found in the examples which are most easily diagonalizable on a computer: $N= 1$ or $2$ in the $S$ representation; $N=3$ in the $A$ representation; and $N=2$ in the $M$ representation. Even though the number of singlets grows extremely quickly with $N$ -- as is to be expected in a tensor model \cite{Geloun:2013kta, Choudhury:2017tax, deMelloKoch:2017bvv, Beccaria:2017aqc, Klebanov:2018nfp} -- we remark that this number remains rather reasonable in the symmetric $\Sp(3)$ model: we find $\sim 10^5$ states, which remains small compared to the $\sim 6.10^8$ states found in $\mO(6)$ tensor quantum mechanics \cite{Klebanov:2018nfp}. As a result, the $\Sp(3)$ model could perhaps provide a more tractable example for future investigations of tensor models at small $N$.   

\begin{table}[h]
\centering
\begin{tabular}{|c||c|c|c||c|c|c|}
\hline
$N$ & $n^{(S)}$ & $d_\cF^{(S)}$ & $I_N^{(S)}$ &  $n^{(A)}$ & $d_\cF^{(A)}$ & $I_N^{(A)}$  \\\hline\hline
1 & 4 & 16 & ${\mathbf 3}$ & -- & -- & --  \\\hline
2 & 20 & 1048576 & ${\mathbf{39}}$ & -- & -- & --  \\\hline
3 &  56 & 72057594037927936 & ${\mathbf{170640}}$ & 14 & 16384 & ${\mathbf 8}$  \\\hline
4 &  120 & $\sim 10^{36}$ & $\sim 10^{14}$ & 48 & 281474976710656 & ${\mathbf{370}}$  \\\hline
\end{tabular}

\vspace{.5cm}

\begin{tabular}{|c||c|c|c|}\hline
N & $n^{(M)}$ & $d_\cF^{(M)}$ & $I_N^{(M)}$\\\hline\hline 
1 & -- & -- & -- \\\hline
2 & 16 & 65536 & ${\mathbf{18}}$ \\\hline
3 & 64 & 18446744073709551616 & ${\mathbf{169826605}}$ \\\hline
4  & 160 & $\sim 10^{48}$  & $\sim 10^{26}$ \\\hline
\end{tabular}
\caption{Summary of state enumaration for small values of $N$.}\label{tab:counting}
\end{table}

\section{Explicit diagonalization at small $N$}\label{sec:diago}

We conclude with a numerical investigation of the energy spectrum at small $N$. We restrict our attention to the two examples which are most easily solvable on a standard laptop: the symmetric model at $N=1$, and the antisymmetric traceless model at $N=3$. Even though both of these systems are too crude to exhibit large $N$ effects, they make up for a good warm-up exercise and provide non-trivial checks of our enumeration results. We leave more in-depth studies in the line of \cite{Klebanov:2018nfp, Pakrouski:2018jcc, Krishnan:2016bvg, Krishnan:2017ztz, Krishnan:2017txw, Krishnan:2018hhu} for future work.

For convenience, we will center the energy spectrum around $0$. It is also possible to shift the Hamiltonian by a term proportional to the charge $\cQ$, so as to make it invariant under the involution $\psi_{abc} \leftrightarrow \psib_{abc}$; as a result, the spectrum will be invariant under $\cQ \leftrightarrow - \cQ$ \footnote{We thank Igor Klebanov for pointing this out.}. We therefore consider a modified version of the original Hamiltonian \eqref{eq:hamiltonian}, of the form
\beq
\tilde{H} = H + x \one + y \cQ \,. 
\eeq
Taking the trace of this equation, we deduce the value of $x$ that makes $\tilde{H}$ traceless:
\beq
x = 
\frac{g}{4} \epsilon_{bg} \epsilon_{dh} \bP_{abc,ade} \bP_{fge,fhc}  =
\left\{ \begin{aligned} 
& \frac{(N+1)^2 (2N+1) N}{18} g &\quad (S)
\\
& \frac{(2N^2+3N+1)^2 (N-2)^2 N}{18 (N-1)}  g &\quad (A)
\end{aligned}\right.
\eeq
To determine the value of $y$ which makes the spectrum invariant under $\cQ \leftrightarrow - \cQ$, we need to solve the equation $H'- H = 2 y Q$, where $H'$ is the Hamiltonian obtained from $H$ by exchanging the roles of $\psi_{abc}$ and $\psib_{abc}$. Using the commutation relations, it is not difficult to show that there exists a solution, given by:
\beq
y = \frac{g}{n} \epsilon_{bg} \epsilon_{dh} \bP_{abc,ade} \bP_{fge,fhc} = \frac{4}{n} x = \left\{ \begin{aligned} 
& \frac{N+1}{3} g &\quad (S)
\\
& \frac{(N-2) (N + 1)}{3 (N-1)} g &\quad (A)
\end{aligned}\right.
\eeq

In practice, we may build up $\tilde{H}$ from any concrete realization of the $\mO(2n)$ Clifford algebra ($n=\Tr\,\bP$):
\beq
\{ \gamma_i , \gamma_j \} = 2 \delta_{ij}\,, \quad i,j = 1, \, \ldots , \, n\,.
\eeq
As for the latter, it is for instance convenient to use a numerical representation in terms of matrices of size $2^n$:
\begin{align}\label{eq:real-gamma}
\gamma_1 &= \sigma_1 \otimes \one \otimes \cdots \otimes \one \nn \\
\gamma_2 &= \sigma_2 \otimes \one \otimes \cdots \otimes \one \nn \\
\gamma_3 &= \sigma_3 \otimes \sigma_1 \otimes \one \otimes \cdots \otimes \one \\
\gamma_4 &= \sigma_3 \otimes \sigma_2 \otimes \one \otimes \cdots \otimes \one \nn \\
&\vdots \nn \\
\gamma_{2n} &= \sigma_3 \otimes \sigma_3 \otimes \cdots \otimes \sigma_2 \nn 
\end{align}
 
\subsection{$N=1$ symmetric model}

\begin{figure}[htb]
\centering
\includegraphics[scale=.5]{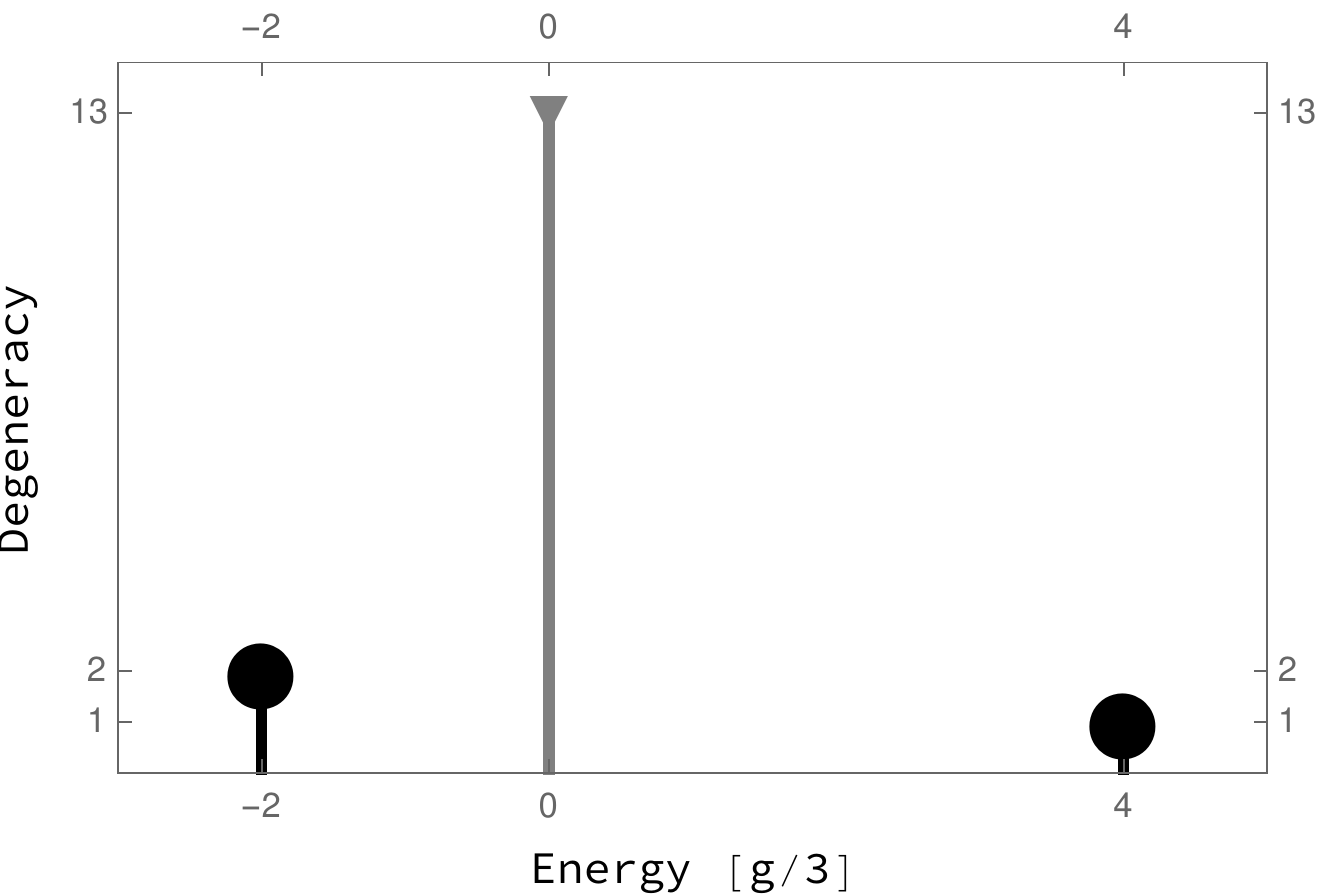}
\caption{The spectrum of $\tilde{H}$ in the symmetric representation with $N=1$. The three $\Sp(N)$ singlets are represented in black, and all the other states have vanishing energy.}\label{fig:spectrum_S1}
\end{figure}

The dimension of the tensor representation being $n=4$, the Hamiltonian $\tilde{H} = H + \frac{2}{3} g ( \one + \cQ )$ can be realized as a $16\times16$ Hermitian matrix. To this purpose, we identify the algebra \eqref{algebra} as
\begin{align}
\Gamma_{111} &= \frac{1}{2} \left( \gamma_1 + i \gamma_5 \right)\,, \nn \\
\Gamma_{112} &= \Gamma_{121} = \Gamma_{211} =  \frac{1}{2\sqrt{3}} \left( \gamma_2 + i \gamma_6 \right) \,, \nn \\
\Gamma_{122} &= \Gamma_{212} = \Gamma_{221} = \frac{1}{2\sqrt{3}} \left( \gamma_3 + i \gamma_7 \right) \,, \nn \\
\Gamma_{222} &= \frac{1}{2} \left(  \gamma_4 + i \gamma_8 \right) \,, \nn  
\end{align}
and reexpress the Hamiltonian as
\begin{align}
\tilde{H} &= \frac{g}{12} ( \gamma_1 \gamma_2 \gamma_3 \gamma_4 -    2 \gamma_1 \gamma_2 \gamma_5 \gamma_6 + 
    \gamma_1 \gamma_2 \gamma_7 \gamma_8 - 
    2 \gamma_1 \gamma_3 \gamma_5 \gamma_7 - 
    \gamma_1 \gamma_3 \gamma_6 \gamma_8 ) \\ 
   & + \frac{g}{12}  ( - 
    \gamma_1 \gamma_4 \gamma_6 \gamma_7 - 
    \gamma_2 \gamma_3 \gamma_5 \gamma_8 - 
    \gamma_2 \gamma_4 \gamma_5 \gamma_7 - 
    2 \gamma_2 \gamma_4 \gamma_6 \gamma_8 + 
    \gamma_3 \gamma_4 \gamma_5 \gamma_6 - 
    2 \gamma_3 \gamma_4 \gamma_7 \gamma_8 + 
    \gamma_5 \gamma_6 \gamma_7 \gamma_8 ) \,. \nn
\end{align}
This matrix can be straightforwardly diagonalized on a computer, and the resulting spectrum is shown on Figure~\ref{fig:spectrum_S1}. Furthermore, the Casimir operator \eqref{eq:casimir} can be used to identify the $\Sp(1)$ singlets: in agreement with the enumeration result \eqref{enumS_r}, we find exactly three such states, with energies $- 2 g$, $\frac{2g}{3}$ and $\frac{4g}{3}$.
They also assume distinct values of the $\U(1)$ charge $\cQ$, which we summarize in the following table:
\begin{center}
\begin{tabular}{|c|c|}\hline
Energy $\tilde{H}$ (in units of $g/3$) & $\U(1)$ charge $\cQ$ \\\hline \hline
$-2.$ &  $2$ \\
 & $-2$ \\\hline
$4.$ &  $0$ \\\hline
\end{tabular}
\end{center} 

\subsection{$N=3$ antisymmetric traceless model}

The traceless Hamiltonian is $\tilde{H} = H + \frac{g}{3}  ( 7 \one + 2 \cQ )$. Following the same method as in the symmetric case, one can represent the generators of the algebra \eqref{algebra} in terms of $28$ $\gamma$ matrices \eqref{eq:real-gamma} of size $16384$. The main new ingredient in this construction is the traceless condition, which can be straightforwardly implemented by a suitable choice of basis.

\begin{figure}[htb]
\centering
\includegraphics[scale=.5]{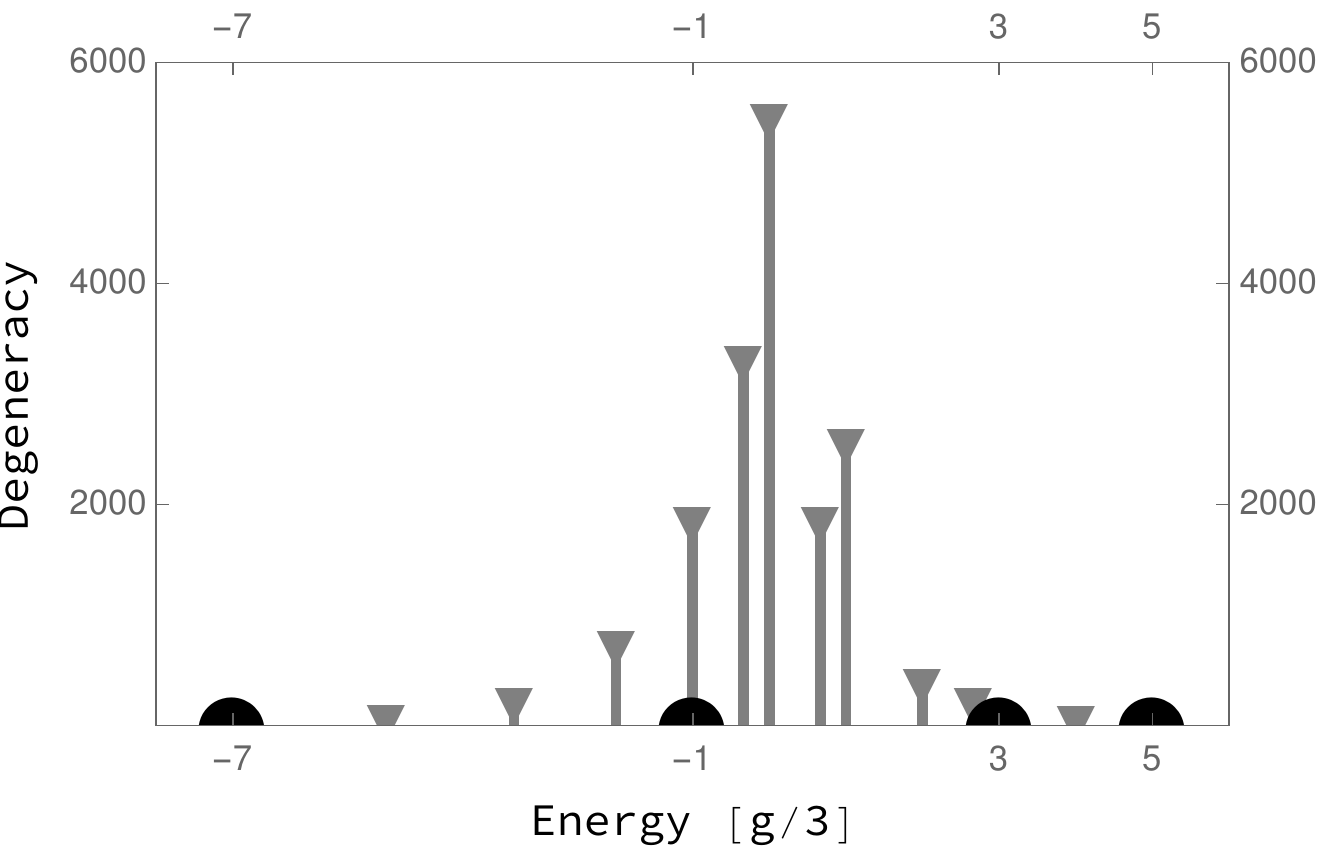} \hspace{1cm}
\includegraphics[scale=.45]{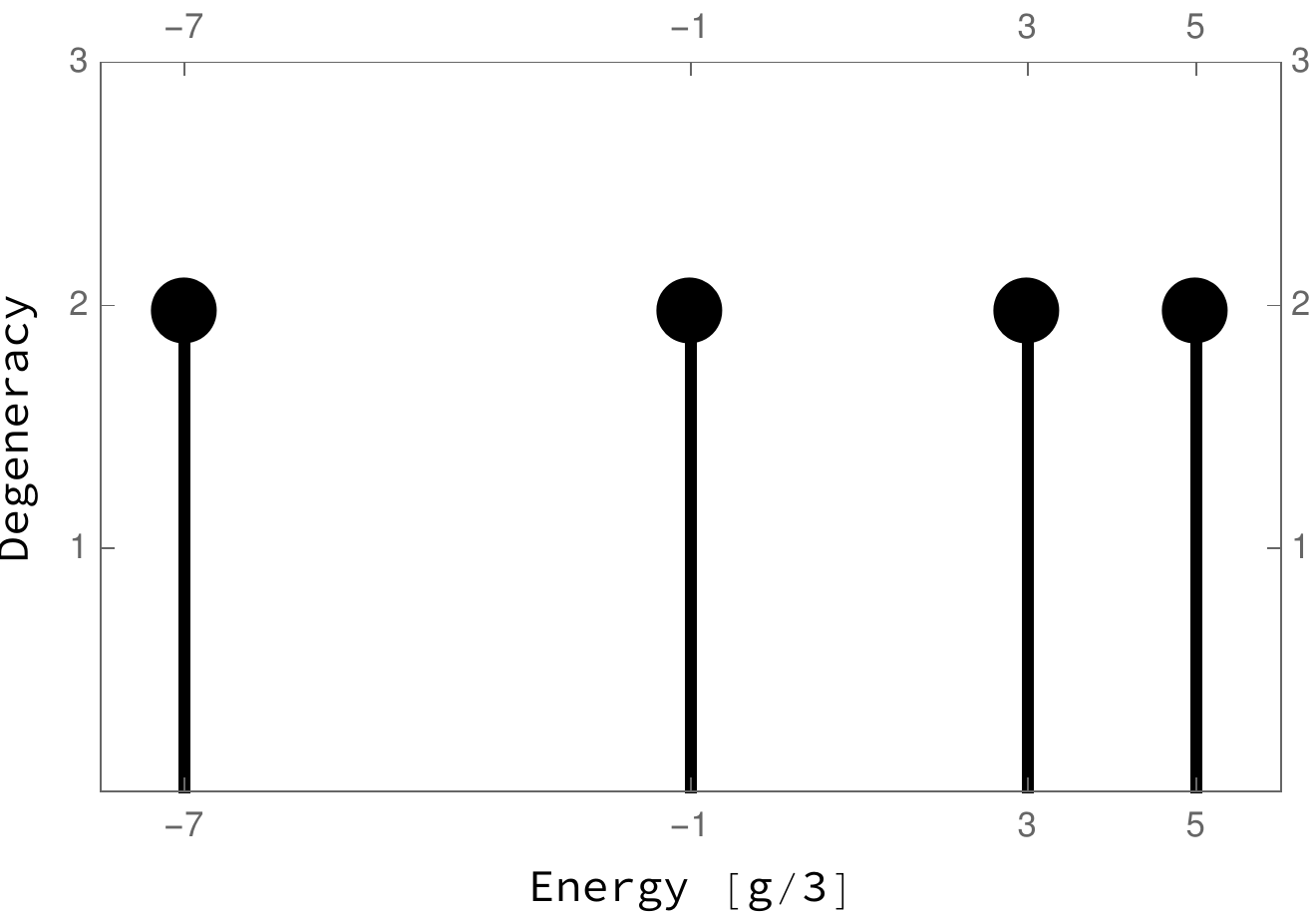}
\caption{The spectrum of $\tilde{H}$ in the antisymmetric traceless representation with $N=3$. The eight $\Sp(N)$ singlets are represented in black, and zoomed in on the right panel.}\label{fig:spectrum_A3}
\end{figure}

We obtain the spectrum plotted in Figure \ref{fig:spectrum_A3}. We find in particular $8$ singlet states, which nicely agrees with the counting \eqref{enumA_r}.
We were also able to determine the charge $\cQ$ of each of the $8$ singlets, which we summarize in the following table: 
\begin{center}
\begin{tabular}{|c|c|}\hline
Energy $\tilde{H}$ (in units of $g/3$) & $\U(1)$ charge $\cQ$ \\\hline \hline
$-7.$ &  $7$ \\
 & $-7$ \\\hline
$-1.$ &  $5$ \\
 & $-5$ \\\hline
$3.$ &  $3$ \\
 & $-3$ \\\hline
$5.$ & $1$ \\
 & $-1$ \\\hline
\end{tabular}
\end{center}
We note that the degeneracy of the four energy doublets is lifted by the charge. Furthermore, it turns out that all the singlets have distinct charges in this simple model.

\section{Conclusion}

In this paper, we have extended the scope of the melonic large $N$ expansion to $\Sp(N)$ rank-$3$ tensors. Analogously to what happens with $\mO(N)$ (anti)symmetrized tensor models \cite{Klebanov:2017nlk, Benedetti:2017qxl, Carrozza:2018ewt}, the existence of the expansion is guaranteed provided that one works with irreducible tensors. In rank three, this leaves us with three inequivalent choices of representations: completely symmetric tensors, completely antisymmetric traceless tensors, and traceless tensors with mixed symmetry (where the trace operation refers to the skew-symmetric matrix $\epsilon$ \eqref{matrixJ}).

Interestingly, and in contrast to $\mO(N)$, the $\Sp(N)$ symmetry makes it possible to write down a non-vanishing tetrahdral interaction for fermions in dimension $1$. This has allowed us to construct three examples of tensor quantum mechanics exhibiting non-trivial SYK-like properties at strong coupling, and without the need to introduce multiple flavours of the gauge group. More precisely, these models can be seen as tensor analogues of the complex SYK model \cite{Sachdev:2015efa}. 

Having enumerated the singlet states and explicitly diagonalized the simplest small $N$ realizations of our models, we hope to have prepared the ground for further numerical investigations, which we leave for future work. 

\section*{Acknowledgements}

\noindent SC would like to thank Igor Klebanov for useful comments on a first version of this manuscript, and Sebastian Steinhaus for helpful advices on numerical integration methods.  

\noindent VP would like to thank Perimeter Institute for hosting him and giving him the chance to work on this project.

\noindent This research was supported in part by Perimeter Institute for Theoretical Physics. Research at Perimeter Institute is supported by the Government of Canada 
through the Department of Innovation, Science and Economic Development Canada and by the Province of Ontario through the Ministry of Research, Innovation and Science.

\appendix

\section{Conventions and useful facts about $\Sp(N)$}\label{sec:facts-spn}

\subsection*{Definition}

The group $\Sp(N)=\U(2N) \cap \Sp(2n , \mathbb{C})$ can be realized as the space of matrices $U \in \mathrm{M}_{2N}(\mathbb{C})$ such that:
\beq
U^{\dagger}U = UU^{\dagger} = \one \qquad \mathrm{and} \qquad U^{\top} \epsilon U = \epsilon  \,,
\eeq 
where $\epsilon$ is an invertible skew-symmetric matrix:
\beq
\epsilon^{\top} = - \epsilon = \epsilon^{-1}\,.
\eeq  
In particular $\epsilon^2 = - \one$, so that $\epsilon$ defines a complex structure. Without loss of generality, we will represent $\epsilon$ by the $2N \times 2N$ block-diagonal matrix
\beq\label{matrixJ}
\epsilon =
	\begin{pmatrix}
		J &  &  \\
		 &  \ddots &  \\
		&  &  J
	\end{pmatrix} \,, \qquad J := \begin{pmatrix} 0 & 1 \\ -1 & 0 \end{pmatrix}\,.
\eeq

$\Sp(N)$ is a compact and simply connected group of (real) dimension $N(2N+1)$. To emphasize the distinction with the symplectic group $\Sp(2N,\mathbb{C})$, which is non-compact, it is sometimes called \emph{compact symplectic group} and denoted $\mathrm{USp}(2N)$. $\Sp(N)$ can also be interpreted as the \emph{quaternionic unitary group} and is therefore quite analogous to $\mO(N)$ (the real unitary group) and $\U(N)$ (the complex unitary group).

\subsection*{Lie algebra and generators}\label{ap:generators}

A $2N \times 2N$ complex matrix $A$ lies in the Lie algebra $\usp(N)$ if and only if:
\beq\label{eq:def-alg-sp}
A^\dagger = - A \qquad \mathrm{and} \qquad A^{\top} \epsilon = - \epsilon A \,. 
\eeq
With our choice of representation of $\epsilon$, it is convenient to write such an $A$ as an $N \times N$ matrix of $2 \times 2$ blocks $A_{kl}$:
\beq
A = \begin{pmatrix}
		A_{11} & \dots & A_{1N} \\
		\vdots &  \ddots & \vdots \\
		A_{N1} & \dots &  A_{NN}
	\end{pmatrix}\,.
\eeq
In view of the relation between $\Sp(N)$ and the quaternions, each block $A_{kl}$ can advantageously be parametrized as:
\beq
A_{kl} =  x_{kl} \one_2 + i \vec{x}_{kl} \cdot \vec{\sigma} \,,
\eeq
where $\vec{\sigma} = (\sigma_1 , \sigma_2 , \sigma_3)$ is a vector of Pauli matrices\footnote{For definiteness: \beq \sigma_1 =\begin{pmatrix} 0 & 1 \\ 1 & 0 \end{pmatrix} \,, \qquad \sigma_2 = \begin{pmatrix} 0 & - i \\ i & 0 \end{pmatrix} \,, \qquad \mathrm{and} \qquad \sigma_3 = \begin{pmatrix} 1 & 0 \\ 0 & -1 \end{pmatrix} \,.\eeq}. The conditions \eqref{eq:def-alg-sp} are then equivalent to
\begin{align}
( x_{kl} , \vec{x}_{kl} ) \in \mathbb{R}^4 \,, \qquad x_{kl} = - x_{lk} \qquad \mathrm{and} \qquad \vec{x}_{kl} = \vec{x}_{lk}\,. 
\end{align} 
At this stage it is convenient to introduce the elementary matrices $E_{k,l}$, defined as having a one in position $(k,l)$ and zeroes elsewhere: 
\beq
(E_{k,l})_{mn} = \delta_{k m} \delta_{l n} \,. 
\eeq
The parametrization of $\usp(N)$ we have just described leads to the following basis of generators: 
\begin{align}
I_{k,l} &: = \left( E_{2k-1 , 2l-1} + E_{2k, 2l} \right) -  \left( E_{2l-1 , 2k-1} + E_{2l , 2k}   \right) \,, & \\
\Sigma^{(1)}_{k,l} &:= i \left( E_{2k-1 , 2l} + E_{2k, 2l-1} \right) + i \left( E_{2l-1 , 2k} + E_{2l , 2k-1}   \right) \,, &\\
\Sigma^{(2)}_{k,l} &:= \left( E_{2k-1 , 2l} - E_{2k, 2l-1} \right) +  \left( E_{2l-1 , 2k} - E_{2l , 2k-1}   \right) \,, & \\
\Sigma^{(3)}_{k,l} &:= i \left( E_{2k-1 , 2l-1} - E_{2k, 2l} \right) + i \left( E_{2l-1 , 2k-1} - E_{2l , 2k}   \right) \,, &\; 1 \leq k < l \leq N\,, \\
\Sigma^{(1)}_m &:=  i \left( E_{2m-1 , 2m} + E_{2m, 2m-1}  \right) \,, & \\
\Sigma^{(2)}_m &:= \left( E_{2m-1 , 2m} - E_{2m, 2m-1} \right) \,, & \\
\Sigma^{(3)}_m &:= i \left( E_{2m-1 , 2m-1} - E_{2m, 2m} \right) \,, &\; 1 \leq m \leq N\,.
\end{align}
In other words $I_{k,l}= (E_{k,l} - E_{l,k})_N \otimes \one_2$, $\Sigma^{(p)}_{k,l}= i (E_{k,l}+E_{l,k})_N \otimes \sigma_p$ and $\Sigma^{(p)}_m = i (E_{m,m})_N \otimes \sigma_p$. It is easy to check that this basis is orthogonal with respect to the inner product $\langle X , Y \rangle = - \tr\left( X Y \right)$, which is itself proportional to the Killing form on $\usp(N)$. Moreover:
\beq
\langle I_{k,l} , I_{k,l} \rangle = \langle \Sigma^{(p)}_{k,l} , \Sigma^{(p)}_{k,l} \rangle = 2 \langle \Sigma^{(p)}_m , \Sigma^{(p)}_m \rangle = 4 \,.  
\eeq

\subsection*{Integration formula for class functions}\label{sec:int-formula}

Any matrix $U \in \Sp(N)$ can be diagonalized by another symplectic matrix $P$:
\beq
U = P D P^{-1}\,, \qquad D= \mathrm{diag}(e^{i \theta_1} , e^{- i \theta_1}, e^{i \theta_2} , e^{- i \theta_2} \ldots ) \,, \qquad P \in \Sp(N)\,,
\eeq
where $\theta_1 \,, \ldots \,, \theta_N$ are real angles. 

Let us denote by $\extd U$ the normalized Haar measure on $\Sp(N)$. Any \emph{class function} $f$ on $\Sp(N)$ reduces to a function of the eigenvalues $f( \theta_1,\,  \ldots ,\, \theta_N):= f( \mathrm{diag}(e^{i \theta_1} , e^{- i \theta_1}, e^{i \theta_2} , e^{- i \theta_2} \ldots) )$. There is then a useful integration formula (see e.g. \cite{simon1996representations}):
\beq\label{int_class}
\int_{\Sp(N)} f(U) \extd U = \int_{[- \pi , \pi]^N} f(\theta_1 , ... , \theta_N ) \, \extd\mu( \theta_1 , \, \ldots , \, \theta_N)\,,
\eeq
where
\beq\label{measure}
\extd\mu( \theta_1 , \, \ldots , \, \theta_N) := \frac{2^{N^2}}{N! (2 \pi)^N} \prod_{i=1}^N \sin^2 \theta_i \prod_{1 \leq i < j \leq N} \left( \cos\theta_i - \cos\theta_j \right)^2 \,.
\eeq

\section{Irreducible $\Sp(N)$ tensors and projectors}\label{app:projectors}

A rank-$3$ complex tensor $T_{a_1 a_2 a_3}$ transforms under $U \in \Sp(N)$ as:
\beq\label{actionSpN}
[U \cdot T]_{a_1 a_2 a_3} = U_{a_1 b_1} U_{a_1 b_1} U_{a_1 b_1} T_{b_1 b_2 b_3}\,.
\eeq
This action commutes with the action of the permutations $\sigma \in \cS_3$:
\beq
[\sigma \triangleright T ]_{a_1 a_2 a_3} := T_{a_{\sigma(1)} a_{\sigma(2)} a_{\sigma(3)}}\,, 
\eeq
and can therefore be decomposed into $\cS_3$ irreducible subrepresentations.
In the language of Young tableaux, one obtains:
\beq\label{eq:young}
\vcenter{\hbox{\begin{Young}  1 \cr \end{Young}}} \otimes \mkern-18mu \vcenter{\hbox{\begin{Young}  2 \cr \end{Young}}} \otimes \mkern-18mu \vcenter{\hbox{\begin{Young}  3 \cr \end{Young}}}  \quad = \sym \quad \oplus \asym \quad \oplus \mixedb \quad \oplus \mixed
\eeq
The first sector is completely symmetric, the second one completely antisymmetric. The last two contain mixed symmetry tensors and yield two equivalent representations; for definiteness, we will only consider mixed tensors of the form $\mixed$. The orthogonal projectors on each of the three representations are:
\begin{align}
\bP^{(S)}_{i_1 i_2 i_3 , j_1 j_2 j_3} &= \frac{1}{3!} \sum_{\sigma \in \mathcal{S}_3} \delta_{i_1j_{\sigma(1)}} \delta_{i_2j_{\sigma(2)}} \delta_{i_3j_{\sigma(3)}} \,, \\
\tilde{\bP}^{(A)}_{i_1 i_2 i_3 , j_1 j_2 j_3} &=  \frac{1}{3!} \sum_{\sigma \in \mathcal{S}_3} (-1)^{\epsilon(\sigma)} \delta_{i_1j_{\sigma(1)}} \delta_{i_2j_{\sigma(2)}} \delta_{i_3j_{\sigma(3)}} \,, \\
\tilde{\bP}^{(M)}_{i_1 i_2 i_3 , j_1 j_2 j_3} &= \frac{1}{3} \left( \delta_{i_1 j_1} \delta_{i_2 j_2} \delta_{i_3 j_3} - \delta_{i_1 j_3} \delta_{i_2 j_2} \delta_{i_3 j_1} \right) \\
&\quad + \frac{1}{6} \left(  \delta_{i_1 j_1} \delta_{i_2 j_3} \delta_{i_3 j_2} +  \delta_{i_1 j_2} \delta_{i_2 j_1} \delta_{i_3 j_3} -  \delta_{i_1 j_2} \delta_{i_2 j_3} \delta_{i_3 j_1} -  \delta_{i_1 j_3} \delta_{i_2 j_1} \delta_{i_3 j_2}\right) \,. \nn
\end{align}

The symmetric representation is already irreducible, the three others are not. To see this, remark that the action \eqref{actionSpN} commutes with the \emph{trace operations}:
\beq
T_{abc} \mapsto \epsilon_{ab} T_{abc} \, , \qquad  T_{abc} \mapsto \epsilon_{ac} T_{abc} \,, \qquad T_{abc} \mapsto \epsilon_{bc} T_{abc} \,.
\eeq
Except for completely symmetric tensors, all sectors contain a non-trivial trace and can therefore be reduced further. To guarantee the existence of a rich large $N$ limit, it is particularly important to remove such vector modes from the representation. This is achieved by acting with the orthogonal projector onto traceless tensors:
\begin{align}
\bP^{(T)}_{i_1 i_2 i_3 , j_1 j_2 j_3} &= \delta_{i_{1}j_{1}} \delta_{i_{2}j_{2}} \delta_{i_{3}j_{3}} \\
& \quad - \frac{2N-1}{(2N-2)(2N+1)} \sum\limits_{p \in \mathcal{A}_3} \epsilon_{i_{p(1)}i_{p(2)}} \epsilon_{j_{p(1)}j_{p(2)}} \delta_{i_{p(3)}j_{p(3)}} \\ \nonumber
	&\quad - \frac{1}{(2N-2)(2N+1)}  \sum\limits_{\substack{ p, q \in \mathcal{A}_3 \\ p \neq q}} \epsilon_{i_{p(1)}i_{p(2)}} \epsilon_{j_{q(1)}j_{q(2)}} \delta_{i_{p(3)}j_{q(3)}}
\end{align}
where $\cA_3 = \{ 1 , (123) , (132) \}$ is the alternating subgroup of $\cS_3$. The operators $\bP^{(S)}$, $\bP^{(A)} := \tilde{\bP}^{(A)}\bP^{(T)}$ and $\bP^{(M)} := \tilde{\bP}^{(M)}\bP^{(T)}$ thus obtained are orthogonal projectors onto irreducible $\Sp(N)$ tensor representations. It is convenient to give these final expressions in graphical form. For instance, $\bP^{(S)}$ can be represented as:
\begin{align}\label{projS}
\bP^{(S)} &= \frac{1}{6} \left( ~\vcenter{\hbox{\includegraphics[width=1cm]{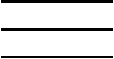}}}~ + ~\vcenter{\hbox{\includegraphics[width=1cm]{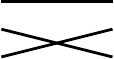}}}~ + ~\vcenter{\hbox{\includegraphics[width=1cm]{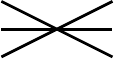}}}~ + ~\vcenter{\hbox{\includegraphics[width=1cm]{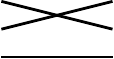}}}~ + ~\vcenter{\hbox{\includegraphics[width=1cm]{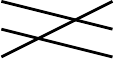}}}~ + ~\vcenter{\hbox{\includegraphics[width=1cm]{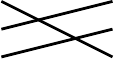}}}~ \right)
\end{align}
Each triplet of strands represents a product of Kronecker deltas, where the indices are ordered from top to bottom on both sides. For the antysimmetric traceless projector we obtain instead: 
\begin{align}\label{projA}
\bP^{(A)} &= \frac{1}{6} \left( ~\vcenter{\hbox{\includegraphics[width=1cm]{Prop1}}}~ - ~\vcenter{\hbox{\includegraphics[width=1cm]{Prop2}}}~ - ~\vcenter{\hbox{\includegraphics[width=1cm]{Prop3}}}~ - ~\vcenter{\hbox{\includegraphics[width=1cm]{Prop4}}}~ + ~\vcenter{\hbox{\includegraphics[width=1cm]{Prop5}}}~ + ~\vcenter{\hbox{\includegraphics[width=1cm]{Prop6}}}~ \right) \nonumber \\
	&-\frac{1}{6(N-1)} \left( ~\vcenter{\hbox{\includegraphics[width=1cm]{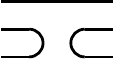}}} ~+~ \vcenter{\hbox{\includegraphics[width=1cm]{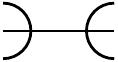}}} ~+~ \vcenter{\hbox{\includegraphics[width=1cm]{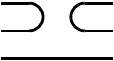}}} ~-~ \vcenter{\hbox{\includegraphics[width=1cm]{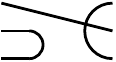}}} ~+~ \vcenter{\hbox{\includegraphics[width=1cm]{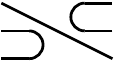}}} ~-~ \vcenter{\hbox{\includegraphics[width=1cm]{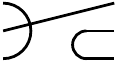}}} \right.  \\
	 & \left. \phantom{-\frac{1}{6(N-1)} \left( \right.} ~-~\vcenter{\hbox{\includegraphics[width=1cm]{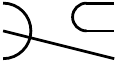}}} ~+~ \vcenter{\hbox{\includegraphics[width=1cm]{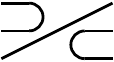}}} ~-~ \vcenter{\hbox{\includegraphics[width=1cm]{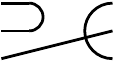}}}~ \right) \nn
\end{align}
The second and third lines correspond to trace removal contributions. Strands connecting two indices on a same side of the drawing encode $\epsilon$ contractions, with the convention that indices are cyclically ordered as $1 \rightarrow 2 \rightarrow 3$: for instance, a strand connecting indices $i_1$ and $i_2$ is associated to a tensor $\epsilon_{i_1 i_2}$, while a strand connecting $j_1$ and $j_3$ encodes a contraction with $\epsilon_{j_3 j_1}= - \epsilon_{j_1 j_3}$.
Finally, the mixed symmetry projector is:
\begin{align}\label{projM}
\bP^{(M)} &= \frac{1}{3} \left( ~\vcenter{\hbox{\includegraphics[width=1cm]{Prop1}}}~ - ~\vcenter{\hbox{\includegraphics[width=1cm]{Prop3}}}~ \right)~ + ~ \frac{1}{6} \left( ~\vcenter{\hbox{\includegraphics[width=1cm]{Prop2}}}~ + ~\vcenter{\hbox{\includegraphics[width=1cm]{Prop4}}}~ - ~\vcenter{\hbox{\includegraphics[width=1cm]{Prop5}}}~ - ~\vcenter{\hbox{\includegraphics[width=1cm]{Prop6}}}~ \right) \nn  \\
	&-\frac{2}{3(2N+1)} ~\vcenter{\hbox{\includegraphics[width=1cm]{Prop8}}} -\frac{1}{3(2N+1)} \left( ~\vcenter{\hbox{\includegraphics[width=1cm]{Prop10}}} ~+~ \vcenter{\hbox{\includegraphics[width=1cm]{Prop12}}} ~+~ \vcenter{\hbox{\includegraphics[width=1cm]{Prop13}}} ~+~ \vcenter{\hbox{\includegraphics[width=1cm]{Prop15}}} \right)  \\
	&-\frac{1}{6(2N+1)} \left( ~\vcenter{\hbox{\includegraphics[width=1cm]{Prop7}}} ~+~ \vcenter{\hbox{\includegraphics[width=1cm]{Prop9}}} ~+~ \vcenter{\hbox{\includegraphics[width=1cm]{Prop11}}} ~+~ \vcenter{\hbox{\includegraphics[width=1cm]{Prop14}}}~ \right) \nn
\end{align}  

\section{Exact expressions of melonic tensor contractions}\label{app:Melons}

$\cM^{(S)}_N$ and $\cM^{(S)}_A$ lead to relatively simple expressions:
\begin{align}
\cM^{(S)}_N & = \frac{\lambda^2}{54}  \frac{2 N^3 + 9 N^2 + 16 N + 9}{N^3}\,,\\
\cM^{(A)}_N & = \frac{\lambda^2}{54}  \frac{2 N^6 - 15 N^5 + 32 N^4 + 21 N^3 - 100 N^2 - 24 N + 48}{N^3 (N-1)^3} \,.
\end{align}
We however obtain much longer expressions in the mixed case:
\begin{align}
\cM^{(M)}_N & = \frac{159 + 684 N + 392 N^2 - 372 N^3 - 220 N^4 + 192 N^5 + 
    128 N^6}{54 N^3 (N-1)^3} \left( {\lambda_1}^2 + {\lambda_2}^2 \right)  \\ 
    &\quad + \frac{-507 - 1557 N - 334 N^2 + 744 N^3 + 296 N^4 + 
    48 N^5 + 32 N^6}{108 N^3 (N-1)^3} \lambda_1 \lambda_2 \,, \nn \\
\cN^{(M)}_N & = \frac{159 + 684 N + 392 N^2 - 372 N^3 - 220 N^4 + 192 N^5 + 
    128 N^6}{54 N^3 (2N+1)^3} {\lambda_1}^2 \nn \\ 
    &\quad + 
 \frac{-507 - 1557 N - 334 N^2 + 744 N^3 + 296 N^4 + 48 N^5 + 
    32 N^6}{108 N^3 (2N+1)^3)} \lambda_1 \lambda_2 \\
    &\quad + \frac{-507 - 1557 N - 334 N^2 + 744 N^3 + 296 N^4 + 
    48 N^5 + 32 N^6}{216 N^3 (2N+1)^3} {\lambda_2}^2 \nn \,.
\end{align}

\section{Character formulas}\label{sec:chi-formulas}

We start by explaining how to get \eqref{Isym_final} from \eqref{Isym_init}. Using the shorthand $f_{klm}:=\cos\frac{\tilde\theta_k + \tilde\theta_l  + \tilde\theta_m }{2}$ the integrand of the latter can be decomposed as 
\beq
\prod_{1 \leq k \leq l \leq m \leq 2N} f_{klm} = \left( \prod_{1 \leq k < l < m \leq 2N} f_{klm} \right) \left( \prod_{\substack{ 1 \leq k, l \leq 2N \\ k \neq l}} f_{kll} \right) \left( \prod_{1 \leq k \leq 2N} f_{kkk} \right)\,.
\eeq
Remembering that the $\tilde\theta$'s take value $\pm \theta_1, \, \ldots , \, \pm \theta_N$, each factor can be computed separately, yielding:
\begin{align}
\prod_{1 \leq k < l < m \leq 2N} f_{klm} &= \left( \prod_{k = 1}^N \cos\frac{\theta_k}{2} \right)^{2N-2} \prod_{\substack{1 \leq k < l < m \leq N \\ \varepsilon_k , \, \varepsilon_l , \, \varepsilon_m = \pm 1}} \cos \frac{\varepsilon_k \theta_k + \varepsilon_l \theta_l + \varepsilon_m \theta_m}{2} \nn \\
&= \left( \prod_{k = 1}^N \cos\frac{\theta_k}{2} \right)^{2N-2} \left( \prod_{1 \leq k < l < m \leq N} \cos \frac{\theta_k + \theta_l + \theta_m}{2} \cos \frac{\theta_k + \theta_l - \theta_m}{2} \right. \\
&  \qquad\qquad\qquad\qquad\qquad \left. \phantom{\prod_{1 \leq k < l < m \leq N}}   \cos \frac{\theta_k - \theta_l + \theta_m}{2} \cos \frac{\theta_k - \theta_l - \theta_m}{2} \right)^2\,, \nn
\end{align}
\begin{align}
 \prod_{\substack{ 1 \leq k, l \leq 2N \\ k \neq l}} f_{kll} &= \left( \prod_{k = 1}^N \cos\frac{\theta_k}{2} \right)^{2} \left( \prod_{\substack{1 \leq k < l \leq N \\ \varepsilon_k , \, \varepsilon_l  = \pm 1}} \cos \frac{2 \varepsilon_k \theta_k + \varepsilon_l \theta_l }{2} \cos \frac{\varepsilon_k \theta_k + 2 \varepsilon_l \theta_l }{2} \right) \nn \\
 &= \left( \prod_{k = 1}^N \cos\frac{\theta_k}{2} \right)^{2} \left( \prod_{1 \leq k < l \leq N } \cos \frac{2 \theta_k + \theta_l }{2} \cos \frac{ \theta_k + 2 \theta_l }{2} \cos \frac{2 \theta_k - \theta_l }{2} \cos \frac{ \theta_k - 2 \theta_l }{2} \right)^2 \,,
\end{align}
and
\beq
\prod_{1 \leq k \leq 2N} f_{kkk} = \left(\prod_{k = 1}^N \cos \frac{3 \theta_k}{2}\right)^2 \,. 
\eeq
All in all, we obtain the desired integrand:
\begin{align}
\prod_{1 \leq k \leq l \leq m \leq 2N} f_{klm} &= \left(\prod_{k = 1}^N \cos \frac{3 \theta_k}{2}\right)^2 \left( \prod_{k = 1}^N \cos\frac{\theta_k}{2} \right)^{2N} \nn \\
&\;  \left( \prod_{1 \leq k < l \leq N } \cos \frac{2 \theta_k + \theta_l }{2} \cos \frac{ \theta_k + 2 \theta_l }{2} \cos \frac{2 \theta_k - \theta_l }{2} \cos \frac{ \theta_k - 2 \theta_l }{2} \right)^2 \\
& \; \left( \prod_{1 \leq k < l < m \leq N} \cos \frac{\theta_k + \theta_l + \theta_m}{2} \cos \frac{\theta_k + \theta_l - \theta_m}{2}  \cos \frac{\theta_k - \theta_l + \theta_m}{2} \cos \frac{\theta_k - \theta_l - \theta_m}{2} \right)^2 . \nn
\end{align}

Formulas \eqref{Iasym_final} and \eqref{Im_final} follow similarly from \eqref{Iasym_init} and \eqref{Im_init}; in the mixed case one may use the factorization:
\begin{align}
\prod_{1 \leq k , l , m \leq 2N} f_{klm} = \left( \prod_{1 \leq k < l < m \leq 2N} f_{klm} \right)^6 \left( \prod_{\substack{ 1 \leq k, l \leq 2N \\ k \neq l}} f_{kll} \right)^3 \left( \prod_{1 \leq k \leq 2N} f_{kkk} \right)\,.
\end{align}

\let\oldbibliography\thebibliography
\renewcommand{\thebibliography}[1]{\oldbibliography{#1}
\setlength{\itemsep}{-1pt}}
\bibliographystyle{JHEP}
\bibliography{Refs.bib}

\providecommand{\href}[2]{#2}\begingroup\raggedright\begin{thebibliography}{10}

\bibitem{expansion1}
R.~Gurau, \emph{{The {$1/N$} expansion of colored tensor models}},
  \href{http://dx.doi.org/10.1007/s00023-011-0101-8}{\emph{Ann. H. Poincar\'e}
  {\bf 12} (2011) 829--847}, [\href{https://arxiv.org/abs/1011.2726}{{\tt
  1011.2726}}].

\bibitem{expansion2}
R.~Gurau and V.~Rivasseau, \emph{{The {$1/N$} expansion of colored tensor
  models in arbitrary dimension}},
  \href{http://dx.doi.org/10.1209/0295-5075/95/50004}{\emph{Europhys. Lett.}
  {\bf 95} (2011) 50004}, [\href{https://arxiv.org/abs/1101.4182}{{\tt
  1101.4182}}].

\bibitem{expansion3}
R.~Gurau, \emph{{The complete {$1/N$} expansion of colored tensor models in
  arbitrary dimension}},
  \href{http://dx.doi.org/10.1007/s00023-011-0118-z}{\emph{Ann. H. Poincar\'e}
  {\bf 13} (2012) 399--423}, [\href{https://arxiv.org/abs/1102.5759}{{\tt
  1102.5759}}].

\bibitem{critical}
V.~Bonzom, R.~Gurau, A.~Riello and V.~Rivasseau, \emph{{Critical behavior of
  colored tensor models in the large {$N$} limit}},
  \href{http://dx.doi.org/10.1016/j.nuclphysb.2011.07.022}{\emph{Nucl. Phys.}
  {\bf B853} (2011) 174--195}, [\href{https://arxiv.org/abs/1105.3122}{{\tt
  1105.3122}}].

\bibitem{RTM}
R.~Gurau, \emph{{Random Tensors}}.
\newblock Oxford University Press, Oxford, 2016.

\bibitem{uncoloring}
V.~Bonzom, R.~Gurau and V.~Rivasseau, \emph{{Random tensor models in the large
  {$N$} limit: Uncoloring the colored tensor models}},
  \href{http://dx.doi.org/10.1103/PhysRevD.85.084037}{\emph{Phys. Rev.} {\bf
  D85} (2012) 084037}, [\href{https://arxiv.org/abs/1202.3637}{{\tt
  1202.3637}}].

\bibitem{expansion4}
V.~Bonzom, \emph{{New {$1/N$} expansions in random tensor models}},
  \href{http://dx.doi.org/10.1007/JHEP06(2013)062}{\emph{JHEP} {\bf 1306}
  (2013) 062}, [\href{https://arxiv.org/abs/1211.1657}{{\tt 1211.1657}}].

\bibitem{expansioin5}
S.~Dartois, V.~Rivasseau and A.~Tanasa, \emph{{The {$1/N$} expansion of
  multi-orientable random tensor models}},
  \href{http://dx.doi.org/10.1007/s00023-013-0262-8}{\emph{Ann. H. Poincar\'e}
  {\bf 15} (2014) 965--984}, [\href{https://arxiv.org/abs/1301.1535}{{\tt
  1301.1535}}].

\bibitem{expansioin6}
R.~Gurau, \emph{{The {$1/N$} expansion of tensor models beyond perturbation
  theory}}, \href{http://dx.doi.org/10.1007/s00220-014-1907-2}{\emph{Commun.
  Math. Phys.} {\bf 330} (2014) 973--1019},
  [\href{https://arxiv.org/abs/1304.2666}{{\tt 1304.2666}}].

\bibitem{Carrozza:2015adg}
S.~Carrozza and A.~Tanasa, \emph{{$O(N)$ Random Tensor Models}},
  \href{http://dx.doi.org/10.1007/s11005-016-0879-x}{\emph{Lett. Math. Phys.}
  {\bf 106} (2016) 1531--1559}, [\href{https://arxiv.org/abs/1512.06718}{{\tt
  1512.06718}}].

\bibitem{thesisLuca}
L.~Lionni, \emph{Colored Discrete Spaces: Higher Dimensional Combinatorial Maps
  and Quantum Gravity}.
\newblock Springer Theses, 2018,
  \href{http://dx.doi.org/10.1007/978-3-319-96023-4}{10.1007/978-3-319-96023-4}.

\bibitem{Bonzom:2018btd}
V.~Bonzom, \emph{{Maximizing the number of edges in three-dimensional colored
  triangulations whose building blocks are balls}},
  \href{https://arxiv.org/abs/1802.06419}{{\tt 1802.06419}}.

\bibitem{universality}
R.~Gurau, \emph{{Universality for Random Tensors}},
  \href{http://dx.doi.org/10.1214/13-AIHP567}{\emph{Ann. Inst. H. Poincare
  Probab. Statist.} {\bf 50} (2014) 1474--1525},
  [\href{https://arxiv.org/abs/1111.0519}{{\tt 1111.0519}}].

\bibitem{melbp}
R.~Gurau and J.~P. Ryan, \emph{{Melons are branched polymers}},
  \href{http://dx.doi.org/10.1007/s00023-013-0291-3}{\emph{Ann. H. Poincar\'e}
  {\bf 15} (2014) 2085--2131}, [\href{https://arxiv.org/abs/1302.4386}{{\tt
  1302.4386}}].

\bibitem{BenGeloun:2011rc}
J.~Ben~Geloun and V.~Rivasseau, \emph{{A renormalizable 4-Dimensional tensor
  field theory}},
  \href{http://dx.doi.org/10.1007/s00220-012-1549-1}{\emph{Commun. Math. Phys.}
  {\bf 318} (2013) 69--109}, [\href{https://arxiv.org/abs/1111.4997}{{\tt
  1111.4997}}].

\bibitem{BenGeloun:2012pu}
J.~Ben~Geloun and D.~O. Samary, \emph{{3D tensor field theory: renormalization
  and one-loop {$\beta$}-functions}},
  \href{http://dx.doi.org/10.1007/s00023-012-0225-5}{\emph{Ann. H. Poincar\'e}
  {\bf 14} (2013) 1599--1642}, [\href{https://arxiv.org/abs/1201.0176}{{\tt
  1201.0176}}].

\bibitem{Samary:2014oya}
D.~Ousmane~Samary, C.~I. Pérez-Sánchez, F.~Vignes-Tourneret and
  R.~Wulkenhaar, \emph{{Correlation functions of a just renormalizable
  tensorial group field theory: the melonic approximation}},
  \href{http://dx.doi.org/10.1088/0264-9381/32/17/175012}{\emph{Class. Quant.
  Grav.} {\bf 32} (2015) 175012}, [\href{https://arxiv.org/abs/1411.7213}{{\tt
  1411.7213}}].

\bibitem{Rivasseau:2017xbk}
V.~Rivasseau and F.~Vignes-Tourneret, \emph{{Constructive tensor field theory:
  The $T^{4}_{4}$ model}},  \href{https://arxiv.org/abs/1703.06510}{{\tt
  1703.06510}}.

\bibitem{Carrozza:2012uv}
S.~Carrozza, D.~Oriti and V.~Rivasseau, \emph{{Renormalization of Tensorial
  Group Field Theories: {Abelian} {$U(1)$} Models in Four Dimensions}},
  \href{http://dx.doi.org/10.1007/s00220-014-1954-8}{\emph{Commun. Math. Phys.}
  {\bf 327} (2014) 603--641}, [\href{https://arxiv.org/abs/1207.6734}{{\tt
  1207.6734}}].

\bibitem{tt2}
S.~Carrozza, D.~Oriti and V.~Rivasseau, \emph{{Renormalization of a {SU(2)}
  Tensorial Group Field Theory in Three Dimensions}},
  \href{http://dx.doi.org/10.1007/s00220-014-1928-x}{\emph{Commun. Math. Phys.}
  {\bf 330} (2014) 581--637}, [\href{https://arxiv.org/abs/1303.6772}{{\tt
  1303.6772}}].

\bibitem{Samary:2012bw}
D.~O. Samary and F.~Vignes-Tourneret, \emph{{Just Renormalizable {TGFT}'s on
  {$U(1)^{d}$} with Gauge Invariance}},
  \href{http://dx.doi.org/10.1007/s00220-014-1930-3}{\emph{Commun. Math. Phys.}
  {\bf 329} (2014) 545--578}, [\href{https://arxiv.org/abs/1211.2618}{{\tt
  1211.2618}}].

\bibitem{thesis}
S.~Carrozza, \emph{Tensorial Methods and Renormalization in Group Field
  Theories}.
\newblock Springer Theses, 2014,
  \href{http://dx.doi.org/10.1007/978-3-319-05867-2}{10.1007/978-3-319-05867-2}.

\bibitem{Carrozza:2014rba}
S.~Carrozza, \emph{{Discrete Renormalization Group for {SU(2)} Tensorial Group
  Field Theory}}, \href{http://dx.doi.org/10.4171/AIHPD/15}{\emph{Ann. Inst.
  Henri Poincar\'e Comb. Phys. Interact.} {\bf 2} (2015) 49--112},
  [\href{https://arxiv.org/abs/1407.4615}{{\tt 1407.4615}}].

\bibitem{IsingD}
V.~Bonzom, R.~Gurau and V.~Rivasseau, \emph{{The {Ising} model on random
  lattices in arbitrary dimensions}},
  \href{http://dx.doi.org/10.1016/j.physletb.2012.03.054}{\emph{Phys. Lett.}
  {\bf B711} (2012) 88--96}, [\href{https://arxiv.org/abs/1108.6269}{{\tt
  1108.6269}}].

\bibitem{bonzom2013universality}
V.~Bonzom, R.~Gurau and M.~Smerlak, \emph{Universality in p-spin glasses with
  correlated disorder}, {\emph{Journal of Statistical Mechanics: Theory and
  Experiment} {\bf 2013} (2013) L02003}.

\bibitem{Delepouve:2015nia}
T.~Delepouve and R.~Gurau, \emph{{Phase Transition in Tensor Models}},
  \href{http://dx.doi.org/10.1007/JHEP06(2015)178}{\emph{JHEP} {\bf 06} (2015)
  178}, [\href{https://arxiv.org/abs/1504.05745}{{\tt 1504.05745}}].

\bibitem{Benedetti:2014qsa}
D.~Benedetti, J.~Ben~Geloun and D.~Oriti, \emph{{Functional Renormalisation
  Group Approach for Tensorial Group Field Theory: a Rank-3 Model}},
  \href{http://dx.doi.org/10.1007/JHEP03(2015)084}{\emph{JHEP} {\bf 03} (2015)
  084}, [\href{https://arxiv.org/abs/1411.3180}{{\tt 1411.3180}}].

\bibitem{Benedetti:2015yaa}
D.~Benedetti and V.~Lahoche, \emph{{Functional Renormalization Group Approach
  for Tensorial Group Field Theory: A Rank-6 Model with Closure Constraint}},
  \href{http://dx.doi.org/10.1088/0264-9381/33/9/095003}{\emph{Class. Quant.
  Grav.} {\bf 33} (2016) 095003}, [\href{https://arxiv.org/abs/1508.06384}{{\tt
  1508.06384}}].

\bibitem{Eichhorn:2017xhy}
A.~Eichhorn and T.~Koslowski, \emph{{Flowing to the continuum in discrete
  tensor models for quantum gravity}},
  \href{http://dx.doi.org/10.4171/AIHPD/52}{\emph{Ann. Inst. H. Poincare Comb.
  Phys. Interact.} {\bf 5} (2018) 173--210},
  [\href{https://arxiv.org/abs/1701.03029}{{\tt 1701.03029}}].

\bibitem{color}
R.~Gurau, \emph{{Colored Group Field Theory}},
  \href{http://dx.doi.org/10.1007/s00220-011-1226-9}{\emph{Commun. Math. Phys.}
  {\bf 304} (2011) 69--93}, [\href{https://arxiv.org/abs/0907.2582}{{\tt
  0907.2582}}].

\bibitem{lost}
R.~Gurau, \emph{{Lost in translation: topological singularities in group field
  theory}},
  \href{http://dx.doi.org/10.1088/0264-9381/27/23/235023}{\emph{Class. Quant.
  Grav.} {\bf 27} (2010) 235023}, [\href{https://arxiv.org/abs/1006.0714}{{\tt
  1006.0714}}].

\bibitem{review}
R.~Gurau and J.~P. Ryan, \emph{{Colored tensor models - a review}},
  \href{http://dx.doi.org/10.3842/SIGMA.2012.020}{\emph{SIGMA} {\bf 8} (2012)
  020}, [\href{https://arxiv.org/abs/1109.4812}{{\tt 1109.4812}}].

\bibitem{Casali:2017tfh}
M.~R. Casali, P.~Cristofori, S.~Dartois and L.~Grasselli, \emph{{Topology in
  colored tensor models via crystallization theory}},
  \href{http://dx.doi.org/10.1016/j.geomphys.2018.01.001}{\emph{J. Geom. Phys.}
  {\bf 129} (2018) 142--167}, [\href{https://arxiv.org/abs/1704.02800}{{\tt
  1704.02800}}].

\bibitem{Sachdev:1992fk}
S.~Sachdev and J.~Ye, \emph{{Gapless spin fluid ground state in a random,
  quantum Heisenberg magnet}},
  \href{http://dx.doi.org/10.1103/PhysRevLett.70.3339}{\emph{Phys. Rev. Lett.}
  {\bf 70} (1993) 3339}, [\href{https://arxiv.org/abs/cond-mat/9212030}{{\tt
  cond-mat/9212030}}].

\bibitem{Kitaev}
A.~Kitaev, \emph{{A simple model of quantum holography}}, {\emph{KITP strings
  seminar and Entanglement 2015 program (Feb. 12, April 7, and May 27, 2015)}
  }.

\bibitem{Maldacena:2016hyu}
J.~Maldacena and D.~Stanford, \emph{{Remarks on the Sachdev-Ye-Kitaev model}},
  \href{http://dx.doi.org/10.1103/PhysRevD.94.106002}{\emph{Phys. Rev.} {\bf
  D94} (2016) 106002}, [\href{https://arxiv.org/abs/1604.07818}{{\tt
  1604.07818}}].

\bibitem{Polchinski:2016xgd}
J.~Polchinski and V.~Rosenhaus, \emph{{The Spectrum in the Sachdev-Ye-Kitaev
  Model}}, \href{http://dx.doi.org/10.1007/JHEP04(2016)001}{\emph{JHEP} {\bf
  04} (2016) 001}, [\href{https://arxiv.org/abs/1601.06768}{{\tt 1601.06768}}].

\bibitem{Gross:2017aos}
D.~J. Gross and V.~Rosenhaus, \emph{{All point correlation functions in SYK}},
  \href{http://dx.doi.org/10.1007/JHEP12(2017)148}{\emph{JHEP} {\bf 12} (2017)
  148}, [\href{https://arxiv.org/abs/1710.08113}{{\tt 1710.08113}}].

\bibitem{Rosenhaus:2018dtp}
V.~Rosenhaus, \emph{{An introduction to the SYK model}},
  \href{https://arxiv.org/abs/1807.03334}{{\tt 1807.03334}}.

\bibitem{Witten:2016iux}
E.~Witten, \emph{{An SYK-Like Model Without Disorder}},
  \href{https://arxiv.org/abs/1610.09758}{{\tt 1610.09758}}.

\bibitem{Klebanov:2016xxf}
I.~R. Klebanov and G.~Tarnopolsky, \emph{{Uncolored Random Tensors, Melon
  Diagrams, and the SYK Models}},
  \href{http://dx.doi.org/10.1103/PhysRevD.95.046004}{\emph{Phys. Rev.} {\bf
  D95} (2017) 046004}, [\href{https://arxiv.org/abs/1611.08915}{{\tt
  1611.08915}}].

\bibitem{Gurau:2016lzk}
R.~Gurau, \emph{{The complete $1/N$ expansion of a SYK--like tensor model}},
  \href{http://dx.doi.org/10.1016/j.nuclphysb.2017.01.015}{\emph{Nucl. Phys.}
  {\bf B916} (2017) 386--401}, [\href{https://arxiv.org/abs/1611.04032}{{\tt
  1611.04032}}].

\bibitem{Gurau:2017qna}
R.~Gurau, \emph{{The {$\imath \epsilon$} prescription in the SYK model}},
  \href{https://arxiv.org/abs/1705.08581}{{\tt 1705.08581}}.

\bibitem{Bonzom:2017pqs}
V.~Bonzom, L.~Lionni and A.~Tanasa, \emph{{Diagrammatics of a colored SYK model
  and of an SYK-like tensor model, leading and next-to-leading orders}},
  \href{http://dx.doi.org/10.1063/1.4983562}{\emph{J. Math. Phys.} {\bf 58}
  (2017) 052301}, [\href{https://arxiv.org/abs/1702.06944}{{\tt 1702.06944}}].

\bibitem{Bulycheva:2017ilt}
K.~Bulycheva, I.~R. Klebanov, A.~Milekhin and G.~Tarnopolsky, \emph{{Spectra of
  Operators in Large $N$ Tensor Models}},
  \href{http://dx.doi.org/10.1103/PhysRevD.97.026016}{\emph{Phys. Rev.} {\bf
  D97} (2018) 026016}, [\href{https://arxiv.org/abs/1707.09347}{{\tt
  1707.09347}}].

\bibitem{Klebanov:2018nfp}
I.~R. Klebanov, A.~Milekhin, F.~Popov and G.~Tarnopolsky, \emph{{Spectra of
  eigenstates in fermionic tensor quantum mechanics}},
  \href{http://dx.doi.org/10.1103/PhysRevD.97.106023}{\emph{Phys. Rev.} {\bf
  D97} (2018) 106023}, [\href{https://arxiv.org/abs/1802.10263}{{\tt
  1802.10263}}].

\bibitem{Pakrouski:2018jcc}
K.~Pakrouski, I.~R. Klebanov, F.~Popov and G.~Tarnopolsky, \emph{{Spectrum of
  Majorana Quantum Mechanics with $O(4)^3$ Symmetry}},
  \href{https://arxiv.org/abs/1808.07455}{{\tt 1808.07455}}.

\bibitem{Krishnan:2016bvg}
C.~Krishnan, S.~Sanyal and P.~N. Bala~Subramanian, \emph{{Quantum Chaos and
  Holographic Tensor Models}},
  \href{http://dx.doi.org/10.1007/JHEP03(2017)056}{\emph{JHEP} {\bf 03} (2017)
  056}, [\href{https://arxiv.org/abs/1612.06330}{{\tt 1612.06330}}].

\bibitem{Krishnan:2017ztz}
C.~Krishnan, K.~V.~P. Kumar and S.~Sanyal, \emph{{Random Matrices and
  Holographic Tensor Models}},
  \href{http://dx.doi.org/10.1007/JHEP06(2017)036}{\emph{JHEP} {\bf 06} (2017)
  036}, [\href{https://arxiv.org/abs/1703.08155}{{\tt 1703.08155}}].

\bibitem{Krishnan:2017txw}
C.~Krishnan and K.~V.~P. Kumar, \emph{{Towards a Finite-$N$ Hologram}},
  \href{http://dx.doi.org/10.1007/JHEP10(2017)099}{\emph{JHEP} {\bf 10} (2017)
  099}, [\href{https://arxiv.org/abs/1706.05364}{{\tt 1706.05364}}].

\bibitem{Krishnan:2018hhu}
C.~Krishnan and K.~V. Pavan~Kumar, \emph{{Exact Solution of a Strongly Coupled
  Gauge Theory in 0+1 Dimensions}},
  \href{http://dx.doi.org/10.1103/PhysRevLett.120.201603}{\emph{Phys. Rev.
  Lett.} {\bf 120} (2018) 201603},
  [\href{https://arxiv.org/abs/1802.02502}{{\tt 1802.02502}}].

\bibitem{Choudhury:2017tax}
S.~Choudhury, A.~Dey, I.~Halder, L.~Janagal, S.~Minwalla and R.~Poojary,
  \emph{{Notes on melonic $O(N)^{q-1}$ tensor models}},
  \href{http://dx.doi.org/10.1007/JHEP06(2018)094}{\emph{JHEP} {\bf 06} (2018)
  094}, [\href{https://arxiv.org/abs/1707.09352}{{\tt 1707.09352}}].

\bibitem{Benedetti:2018goh}
D.~Benedetti and R.~Gurau, \emph{{2PI effective action for the SYK model and
  tensor field theories}},
  \href{http://dx.doi.org/10.1007/JHEP05(2018)156}{\emph{JHEP} {\bf 05} (2018)
  156}, [\href{https://arxiv.org/abs/1802.05500}{{\tt 1802.05500}}].

\bibitem{Giombi:2017dtl}
S.~Giombi, I.~R. Klebanov and G.~Tarnopolsky, \emph{{Bosonic tensor models at
  large $N$ and small $\epsilon$}},
  \href{http://dx.doi.org/10.1103/PhysRevD.96.106014}{\emph{Phys. Rev.} {\bf
  D96} (2017) 106014}, [\href{https://arxiv.org/abs/1707.03866}{{\tt
  1707.03866}}].

\bibitem{Prakash:2017hwq}
S.~Prakash and R.~Sinha, \emph{{A Complex Fermionic Tensor Model in $d$
  Dimensions}}, \href{http://dx.doi.org/10.1007/JHEP02(2018)086}{\emph{JHEP}
  {\bf 02} (2018) 086}, [\href{https://arxiv.org/abs/1710.09357}{{\tt
  1710.09357}}].

\bibitem{Benedetti:2017fmp}
D.~Benedetti, S.~Carrozza, R.~Gurau and A.~Sfondrini, \emph{{Tensorial
  Gross-Neveu models}},
  \href{http://dx.doi.org/10.1007/JHEP01(2018)003}{\emph{JHEP} {\bf 01} (2018)
  003}, [\href{https://arxiv.org/abs/1710.10253}{{\tt 1710.10253}}].

\bibitem{Giombi:2018qgp}
S.~Giombi, I.~R. Klebanov, F.~Popov, S.~Prakash and G.~Tarnopolsky,
  \emph{{Prismatic Large $N$ Models for Bosonic Tensors}},
  \href{http://dx.doi.org/10.1103/PhysRevD.98.105005}{\emph{Phys. Rev.} {\bf
  D98} (2018) 105005}, [\href{https://arxiv.org/abs/1808.04344}{{\tt
  1808.04344}}].

\bibitem{Ferrari:2017ryl}
F.~Ferrari, \emph{{The Large D Limit of Planar Diagrams}},
  \href{https://arxiv.org/abs/1701.01171}{{\tt 1701.01171}}.

\bibitem{Azeyanagi:2017drg}
T.~Azeyanagi, F.~Ferrari and F.~I. Schaposnik~Massolo, \emph{{Phase Diagram of
  Planar Matrix Quantum Mechanics, Tensor, and Sachdev-Ye-Kitaev Models}},
  \href{http://dx.doi.org/10.1103/PhysRevLett.120.061602}{\emph{Phys. Rev.
  Lett.} {\bf 120} (2018) 061602},
  [\href{https://arxiv.org/abs/1707.03431}{{\tt 1707.03431}}].

\bibitem{Ferrari:2017jgw}
F.~Ferrari, V.~Rivasseau and G.~Valette, \emph{{A New Large N Expansion for
  General Matrix-Tensor Models}},  \href{https://arxiv.org/abs/1709.07366}{{\tt
  1709.07366}}.

\bibitem{Azeyanagi:2017mre}
T.~Azeyanagi, F.~Ferrari, P.~Gregori, L.~Leduc and G.~Valette, \emph{{More on
  the New Large $D$ Limit of Matrix Models}},
  \href{http://dx.doi.org/10.1016/j.aop.2018.04.010}{\emph{Annals Phys.} {\bf
  393} (2018) 308--326}, [\href{https://arxiv.org/abs/1710.07263}{{\tt
  1710.07263}}].

\bibitem{Vasiliev:2018zer}
M.~A. Vasiliev, \emph{{From Coxeter Higher-Spin Theories to Strings and Tensor
  Models}}, \href{http://dx.doi.org/10.1007/JHEP08(2018)051}{\emph{JHEP} {\bf
  08} (2018) 051}, [\href{https://arxiv.org/abs/1804.06520}{{\tt 1804.06520}}].

\bibitem{TASI_tensor-review}
I.~R. Klebanov, F.~Popov and G.~Tarnopolsky, \emph{{TASI Lectures on Large $N$
  Tensor Models}},  in \emph{{Proceedings, Theoretical Advanced Study Institute
  in Elementary Particle Physics: Physics at the Fundamental Frontier (TASI
  2017): Boulder, CO, USA, June 5-30, 2017}}, vol.~TASI2017, p.~004, 2018.
\newblock \href{https://arxiv.org/abs/1808.09434}{{\tt 1808.09434}}.
\newblock \href{http://dx.doi.org/10.22323/1.305.0004}{DOI}.

\bibitem{Bonzom:2018jfo}
V.~Bonzom, V.~Nador and A.~Tanasa, \emph{{Diagrammatic proof of the large $N$
  melonic dominance in the SYK model}},
  \href{https://arxiv.org/abs/1808.10314}{{\tt 1808.10314}}.

\bibitem{Klebanov:2017nlk}
I.~R. Klebanov and G.~Tarnopolsky, \emph{{On Large $N$ Limit of Symmetric
  Traceless Tensor Models}},
  \href{http://dx.doi.org/10.1007/JHEP10(2017)037}{\emph{JHEP} {\bf 10} (2017)
  037}, [\href{https://arxiv.org/abs/1706.00839}{{\tt 1706.00839}}].

\bibitem{Gurau:2017qya}
R.~Gurau, \emph{{The $1/N$ expansion of tensor models with two symmetric
  tensors}}, \href{http://dx.doi.org/10.1007/s00220-017-3055-y}{\emph{Commun.
  Math. Phys.} (2017) }, [\href{https://arxiv.org/abs/1706.05328}{{\tt
  1706.05328}}].

\bibitem{Benedetti:2017qxl}
D.~Benedetti, S.~Carrozza, R.~Gurau and M.~Kolanowski, \emph{{The $1/N$
  expansion of the symmetric traceless and the antisymmetric tensor models in
  rank three}},  \href{https://arxiv.org/abs/1712.00249}{{\tt 1712.00249}}.

\bibitem{Carrozza:2018ewt}
S.~Carrozza, \emph{{Large $N$ limit of irreducible tensor models: $O(N)$
  rank-$3$ tensors with mixed permutation symmetry}},
  \href{http://dx.doi.org/10.1007/JHEP06(2018)039}{\emph{JHEP} {\bf 06} (2018)
  039}, [\href{https://arxiv.org/abs/1803.02496}{{\tt 1803.02496}}].

\bibitem{Sachdev:2015efa}
S.~Sachdev, \emph{{Bekenstein-Hawking Entropy and Strange Metals}},
  \href{http://dx.doi.org/10.1103/PhysRevX.5.041025}{\emph{Phys. Rev.} {\bf X5}
  (2015) 041025}, [\href{https://arxiv.org/abs/1506.05111}{{\tt 1506.05111}}].

\bibitem{Gubser:2017qed}
S.~S. Gubser, M.~Heydeman, C.~Jepsen, S.~Parikh, I.~Saberi, B.~Stoica et~al.,
  \emph{{Signs of the time: Melonic theories over diverse number systems}},
  \href{http://dx.doi.org/10.1103/PhysRevD.98.126007}{\emph{Phys. Rev.} {\bf
  D98} (2018) 126007}, [\href{https://arxiv.org/abs/1707.01087}{{\tt
  1707.01087}}].

\bibitem{Gubser:2018yec}
S.~S. Gubser, C.~Jepsen, Z.~Ji and B.~Trundy, \emph{{Higher melonic theories}},
  \href{http://dx.doi.org/10.1007/JHEP09(2018)049}{\emph{JHEP} {\bf 09} (2018)
  049}, [\href{https://arxiv.org/abs/1806.04800}{{\tt 1806.04800}}].

\bibitem{Gross:2016kjj}
D.~J. Gross and V.~Rosenhaus, \emph{{A Generalization of Sachdev-Ye-Kitaev}},
  \href{http://dx.doi.org/10.1007/JHEP02(2017)093}{\emph{JHEP} {\bf 02} (2017)
  093}, [\href{https://arxiv.org/abs/1610.01569}{{\tt 1610.01569}}].

\bibitem{Kitaev:2017awl}
A.~Kitaev and S.~J. Suh, \emph{{The soft mode in the Sachdev-Ye-Kitaev model
  and its gravity dual}},
  \href{http://dx.doi.org/10.1007/JHEP05(2018)183}{\emph{JHEP} {\bf 05} (2018)
  183}, [\href{https://arxiv.org/abs/1711.08467}{{\tt 1711.08467}}].

\bibitem{Aharony:2003sx}
O.~Aharony, J.~Marsano, S.~Minwalla, K.~Papadodimas and M.~Van~Raamsdonk,
  \emph{{The Hagedorn - deconfinement phase transition in weakly coupled large
  N gauge theories}},
  \href{http://dx.doi.org/10.4310/ATMP.2004.v8.n4.a1}{\emph{Adv. Theor. Math.
  Phys.} {\bf 8} (2004) 603--696},
  [\href{https://arxiv.org/abs/hep-th/0310285}{{\tt hep-th/0310285}}].

\bibitem{Beccaria:2017aqc}
M.~Beccaria and A.~A. Tseytlin, \emph{{Partition function of free conformal
  fields in 3-plet representation}},
  \href{http://dx.doi.org/10.1007/JHEP05(2017)053}{\emph{JHEP} {\bf 05} (2017)
  053}, [\href{https://arxiv.org/abs/1703.04460}{{\tt 1703.04460}}].

\bibitem{Geloun:2013kta}
J.~Ben~Geloun and S.~Ramgoolam, \emph{{Counting tensor model observables and
  branched covers of the 2-Sphere}},
  \href{https://arxiv.org/abs/1307.6490}{{\tt 1307.6490}}.

\bibitem{deMelloKoch:2017bvv}
R.~de~Mello~Koch, R.~Mello~Koch, D.~Gossman and L.~Tribelhorn, \emph{{Gauge
  Invariants, Correlators and Holography in Bosonic and Fermionic Tensor
  Models}}, \href{http://dx.doi.org/10.1007/JHEP09(2017)011}{\emph{JHEP} {\bf
  09} (2017) 011}, [\href{https://arxiv.org/abs/1707.01455}{{\tt 1707.01455}}].

\bibitem{BenGeloun:2017vwn}
J.~Ben~Geloun and S.~Ramgoolam, \emph{{Tensor Models, Kronecker coefficients
  and Permutation Centralizer Algebras}},
  \href{http://dx.doi.org/10.1007/JHEP11(2017)092}{\emph{JHEP} {\bf 11} (2017)
  092}, [\href{https://arxiv.org/abs/1708.03524}{{\tt 1708.03524}}].

\bibitem{simon1996representations}
B.~Simon, \emph{Representations of finite and compact groups}.
\newblock No.~10. American Mathematical Soc., 1996.

\end{thebibliography}\endgroup

\end{document}